\documentclass[12pt]{article}
\usepackage[colorlinks=true, allcolors=blue]{hyperref}
\usepackage[colorinlistoftodos, textsize=tiny]{todonotes}

\usepackage{bm}
\usepackage{subcaption}
\usepackage{diagbox}
\usepackage{times}

\usepackage[english]{babel}
\usepackage[utf8x]{inputenc}
\usepackage[T1]{fontenc}

\usepackage{amsmath,amsfonts,amssymb}
\usepackage{graphicx}

\usepackage{bm}
\usepackage{subcaption}

\presetkeys%
    {todonotes}%
    {noline,backgroundcolor=green!40}{}

\title{A recurrent neural network for classification of unevenly sampled variable stars}
\author{Brett Naul, 
Joshua S. Bloom, 
Fernando P\'{e}rez, 
St\'{e}fan van der Walt
}

\begin{document}
\baselineskip12pt
\maketitle
\begin{abstract}
Astronomical surveys of celestial sources produce  streams of noisy time series measuring flux versus time (``light curves''). Unlike in many other physical domains, however, large (and source-specific) temporal gaps in data arise naturally due to intranight cadence choices as well as diurnal and seasonal constraints~\cite{1996LEVINE,pojmanski2002all,2013murphy,2014Ridgway,2016Djorgovski}. With nightly observations of millions of variable stars and transients from upcoming surveys~\cite{2014Ridgway,Kantor2014LSST}, efficient and accurate discovery and classification techniques on noisy, irregularly sampled data must be employed with minimal human-in-the-loop involvement. Machine learning for inference tasks on such data traditionally requires the laborious hand-coding of domain-specific numerical summaries of raw data (``features'')~\cite{bloom2012review}. Here we present a novel unsupervised autoencoding recurrent neural network~\cite{hinton2006reducing} (RNN) that makes explicit use of sampling times and known heteroskedastic noise properties. When trained on optical variable star catalogs, this network produces supervised classification models that rival other best-in-class approaches. We find that autoencoded features learned on one time-domain survey perform nearly as well when applied to another survey. These networks can continue to learn from new unlabeled observations and may be used in other unsupervised tasks such as forecasting and anomaly detection.
\end{abstract}

The RNN feature extraction architecture proposed (Fig.~\ref{fig:network_auto}) consists of two components: an
encoder, which takes a time series as input and produces a fixed-length feature vector as output, and a
decoder, which translates the feature vector representation back into an output time series.
The principal advantages of our architecture over a standard RNN autoencoder~\cite{hinton2006reducing} are the
eative handling of the sampling times and the explicit use of measurement uncertainty in the
loss function.

Specifically, the autoencoder network is trained with times and measurements as inputs and those same
measurement values as outputs. The mean squared reconstruction error of the output sequence is minimized, using backpropagation and gradient descent. In the case where individual measurement errors are available, the reconstruction error at each time step can be weighted (in analogy with standard weighted least squares regression) to reduce the penalty
for reconstruction errors when the measurement error is large (see Methods).

The feature vector is then taken to be the last element of the output sequence of the last encoder layer, so its dimension is equal to the number of hidden units in that layer. The fixed-length embedding vector produced by the encoder contains sufficient information to approximately reconstruct the input signal, so it may be thought of as a low-dimensional feature representation
of the input data. Although we focus here on an autoencoder model for feature extraction, the decoder portion of the network
can also be trained directly to solve classification or regression problems, as described in detail
in the Supplementary Information (SI).

To investigate the utility of the automatically extracted features, we train an
autoencoder model to reconstruct a set of (unlabeled) light curves and then use the
resulting features to train a classifier to predict
the variable star class. Here we use the 50,124 light curves from the All Sky Automated
Survey (ASAS) Catalog of Variable stars~\cite{pojmanski2002all}.
The autoencoder is trained using the full set of both labeled and unlabeled light curves,
and the resulting features are then used to build a model to solve the supervised
classification task.
We compare the resulting classifier to a model which uses expert-chosen features and
demonstrate that the autoencoding features perform at least as well, and in some cases
better, than the hand-selected features.

Figure~\ref{fig:asas_reconstruct} depicts some examples of reconstructed light curves for
an embedding of length 64 (the other parameters of the autoencoder are described in
Methods); the examples are chosen to represent the 25th and 75th
percentile of reconstruction error in order to show the range of different qualities of
reconstruction. The model is able to effectively represent light curves which exhibit
relatively smooth behavior, even in the presence of large gaps between samples; however,
curves that vary more rapidly (i.e., those with small periods) are less likely to be
reconstructed accurately. As such we also trained an autoencoder model on the
period-folded values (replacing time with phase) using the measured periods~\cite{richards2012construction}. Figure~\ref{fig:asas_reconstruct} shows that
the resulting reconstructions are improved, especially in the case of
the low-period signal. The effect of period on the accuracy of autoencoder reconstructions is explored further using simulated data (see SI). In what follows, we use autoencoders trained on period-folded light curve data.

To evaluate the usefulness of our features for classification, we identify a
subset of ASAS light curves from five classes of variable stars
(see Methods for details). We then train a random forest classifier~\cite{breiman2001random} using the autoencoder-generated features for 80\% of the samples from each class, along with the
means and standard deviations of each light curve which are removed in preprocessing (see
Methods).
The resulting estimator achieves 98.8\%
average accuracy on the validation sets across five 80/20 train/validation splits.

As a baseline, we also constructed random forest classifiers using two sets of standard
features for variable star classification.
The first are the features used in Richards et al.~\cite{richards2011machine} (henceforth
abbreviated ``Richards et al.\ features''); the features are implemented
in the Cesium ML project~\cite{naul2016cesium} and also as part of the FATS
package~\cite{2015Nun}. These features have been used by numerous studies (e.g., refs~\cite{2011Dubath,2014Nun,2015Miller}) including state-of-the-art classification
performance on the ASAS survey and remain competitive against other classification
methods~\cite{2015Kugler}.
The second set of features consists of those used by Kim \& Bailer-Jones~\cite{kim2016package} (henceforth
referred to ``Kim/Bailer-Jones features'') and implemented in the Upsilon package.
Some features are shared between the two, and each set of features is an aggregation of
features from many different works. In both cases we use the same hyperparameter
selection technique described in Figure~\ref{fig:confusion}.

The Richards et al. features achieve the best average validation accuracy at 99.4\% across
the same five splits, as shown in Table~\ref{tab:accuracy}.
However, it is worth noting that the same features were also used in the labeling of the
training set~\cite{richards2012construction}, so it is not surprising that they achieve
almost perfect classification accuracy for this problem.
The Kim/Bailer-Jones features, which may provide a more natural baseline, achieve 98.8\%
validation accuracy, comparable to that of the autoencoder model.

Our second example applies the same feature extraction methodology to variable star light
curves from the Lincoln Near-Earth Asteroid Research (LINEAR)
survey~\cite{sesar2013exploring,palaversa2013exploring}.
The LINEAR dataset consists of 5,204 labeled light curves from five classes of variable
star (see Methods for details).
Unlike in the ASAS example above, here all the available light curves are labeled, so
there is no additional unlabeled data to leverage in order to improve the quality of the
extracted features.
We find that the autoencoder features outperform the Richards et al.\ features
by 0.38\% and the Kim/Bailer-Jones features by 1.61\% (see Table~\ref{tab:accuracy}).
In particular, the autoencoder-based
model correctly classifies all but 1 RR Lyrae, suggesting that perhaps some autoencoder
features could help improve the performance of the Richards et al.\ or Kim/Bailer-Jones
features for that specific discrimination task.

Finally, we followed the same procedure to train an autoencoder and subsequently a random
forest classifier to predict the classes of 21,474 variable stars from the MACHO
catalog~\cite{alcock1996macho}. Once again, our autoencoder approach achieves the best
validation accuracy of the three methods considered, averaging 93.6\% compared to 90.5\%
and 89.0\% for the Richards et al.\ and Kim/Bailer-Jones feature sets, respectively.

As shown, training a autoencoder RNN to represent
time series sequences can produce informative high-level feature representations
in an unsupervised setting. Rather than require fixed-length time series, our approach
naturally accommodates variable length inputs. Other unsupervised techniques for feature
extraction on irregular time-series data have also been developed (e.g., clustering
methods~\cite{mackenzie2016clustering}), but those scale quadradically in the number of
training examples, whereas our approach scales linearly. Moreover, our approach explicitly
accounts for measurement noise. The resulting features are shown to be
comparable or better for supervised classification tasks than
traditional hand-coded features. As new sources accumulate without known labels, the
unsupervised and on-line nature of such networks should ensure continued model
improvements in a way not possible with a fixed number of features.
When metadata is also available (e.g., color and sky position for astronomical variables),
features derived from such non-temporal data can be easily used in conjunction with the
auto-encoded features of the time series.

While the autoencoder approach we have described is well-suited to tasks involving
a relatively large amount of (labeled or unlabeled) data, future research
should study the efficacy of cross-domain
transfer learning, where feature representations learned from a very large dataset are applied
to another problem where fewer examples are available. Another promising application is unsupervised data
exploration, such as clustering; autoencoding features could be used as a generic
lower-dimensional representation like t-SNE~\cite{maaten2008visualizing} to identify
outliers/anomalies in new data.
Autoencoders could also act as non-parametric interpolators tuned to the domain on which
the models are trained. Finally, while we have focused on single-channel time series, the
proposed network is easily extensible to multi-channel time series data as well as
multi-dimensional time series, such as unevenly sampled sequential imaging.

\section*{Acknowledgements}  
We would thank Yann LeCun and Farid El Gabaly for helpful discussions, and Aaron Culich for computational assistance. This work is
supported by the Gordon and Betty Moore Foundation Data-Driven Discovery and NSF BIGDATA
Grant \#1251274. Computation was provided by the Pacific Research Platform program through NSF ACI \#1541349, OCI \#1246396, UCOP,
Calit2, and BRC at UC Berkeley.

\section*{Author information}
\subsection*{Contributions}
B.N.\ implemented and trained the networks, assembled the machine learning results,
and generated the first drafts of the paper and figures.
J.S.B.\ conceived of the project, assembled the astronomical light curves,
and oversaw the supervised training portions.
F.P. provided theoretical input. S.vdW. discussed the results and commented on the manuscript.

\subsection*{Affiliations}
\textit{Department of Astronomy, University of California, Berkeley, CA 94720, USA.} \\
B. Naul, J. S. Bloom \\
\\
\textit{Department of Statistics, Berkeley, CA 94720, USA.} \\
F. P\'{e}rez\\
\\
\textit{Berkeley Institute for Data Science, University of California, Berkeley, CA 94720, USA.} \\
S. van der Walt

\subsection*{Corresponding author}
Correspondence to B. Naul.

\begin{figure}[!htb]
    \centering
    \includegraphics[width=0.75\textwidth]{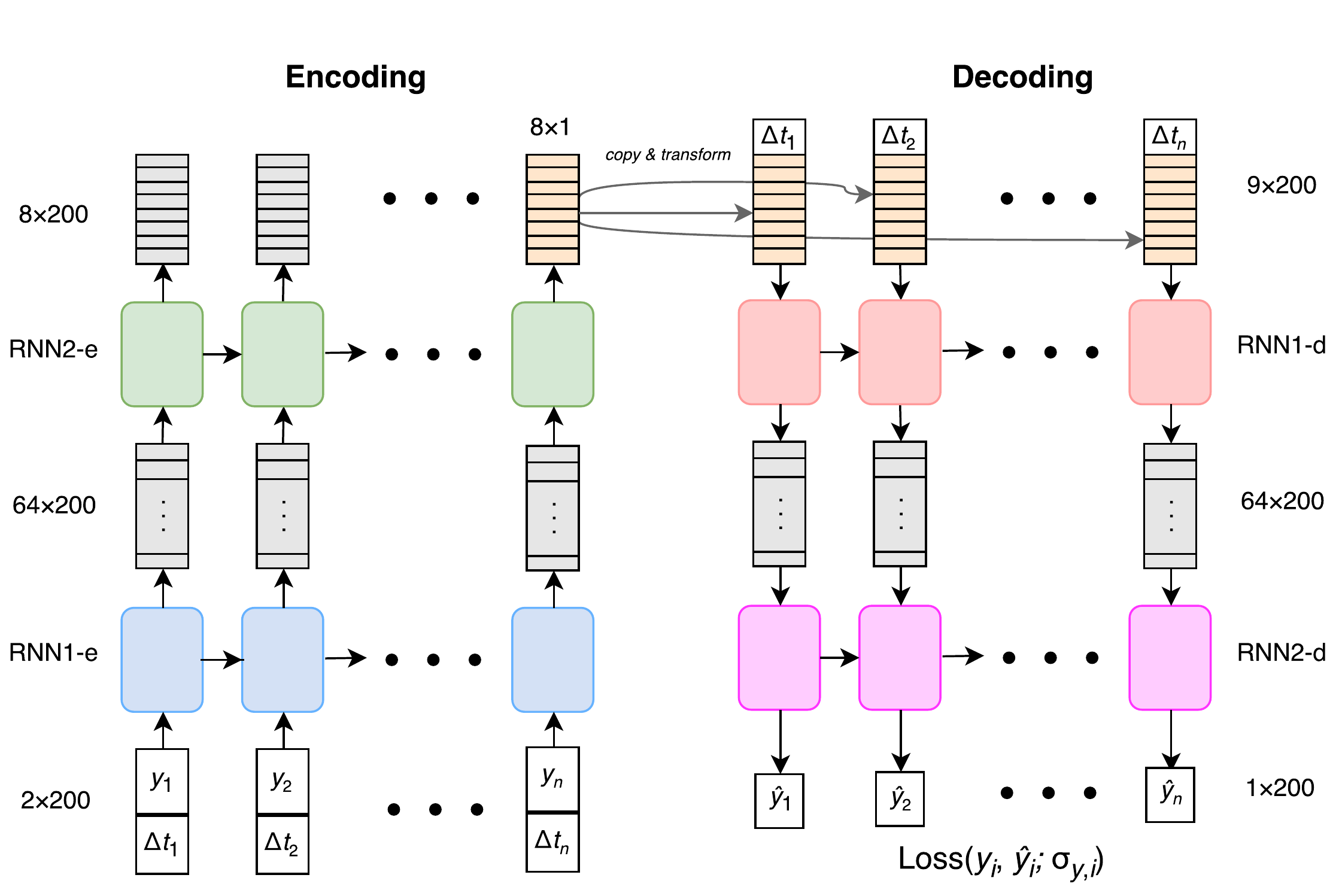}
    \caption{\small \textbf{Diagram of an RNN encoder/decoder architecture for irregularly
        sampled time series data.}
        This network uses two RNN layers (specifically, bidirectional gated recurrent
        units (GRU)~\cite{cho2014learning,schuster1997bidirectional})
        of size 64 for
        encoding and two for decoding, with a feature embedding size of 8. The encoder
        takes as inputs the measurement values as well the sampling times (more
        specifically, the differences between sampling times); the sequence is
        processed by a hidden recurrent layer to produce a new sequence, which can
        then be used as the input to another hidden recurrent layer, etc. The
        fixed-length embedding is constructed by passing the output of the last
        recurrent layer into a single fully-connected layer with linear activation
        function and the desired output size. The decoder first repeats the
        fixed-length embedding $n_T$ times, where $n_T$ is the length of the desired
        output sequence, and then appends the sampling time differences to the
        corresponding elements of the resulting vector sequence.
        The sampling times are passed to both the encoder and decoder; the
        feature vector characterizes the functional form of the signal, but the sampling times
        are needed to determine the points at which that function should be evaluated.
        The remainder of the decoder network is another series of recurrent layers, with a
        final linear layer to generate the output sequence.
        We also apply 25\% dropout~\cite{srivastava2014dropout} between recurrent layers,
        which we omit from the figure for simplicity.
  In our model we
       take the number and size of recurrent layers in the encoder and decoder
        modules to be equal, but in general the two components are entirely distinct
    and need not share any architectural similarities.
    }
    \label{fig:network_auto}
\end{figure}

\begin{figure}[htb!]
    \centering
    \begin{subfigure}[b]{0.495\textwidth}
        \includegraphics[width=\textwidth]{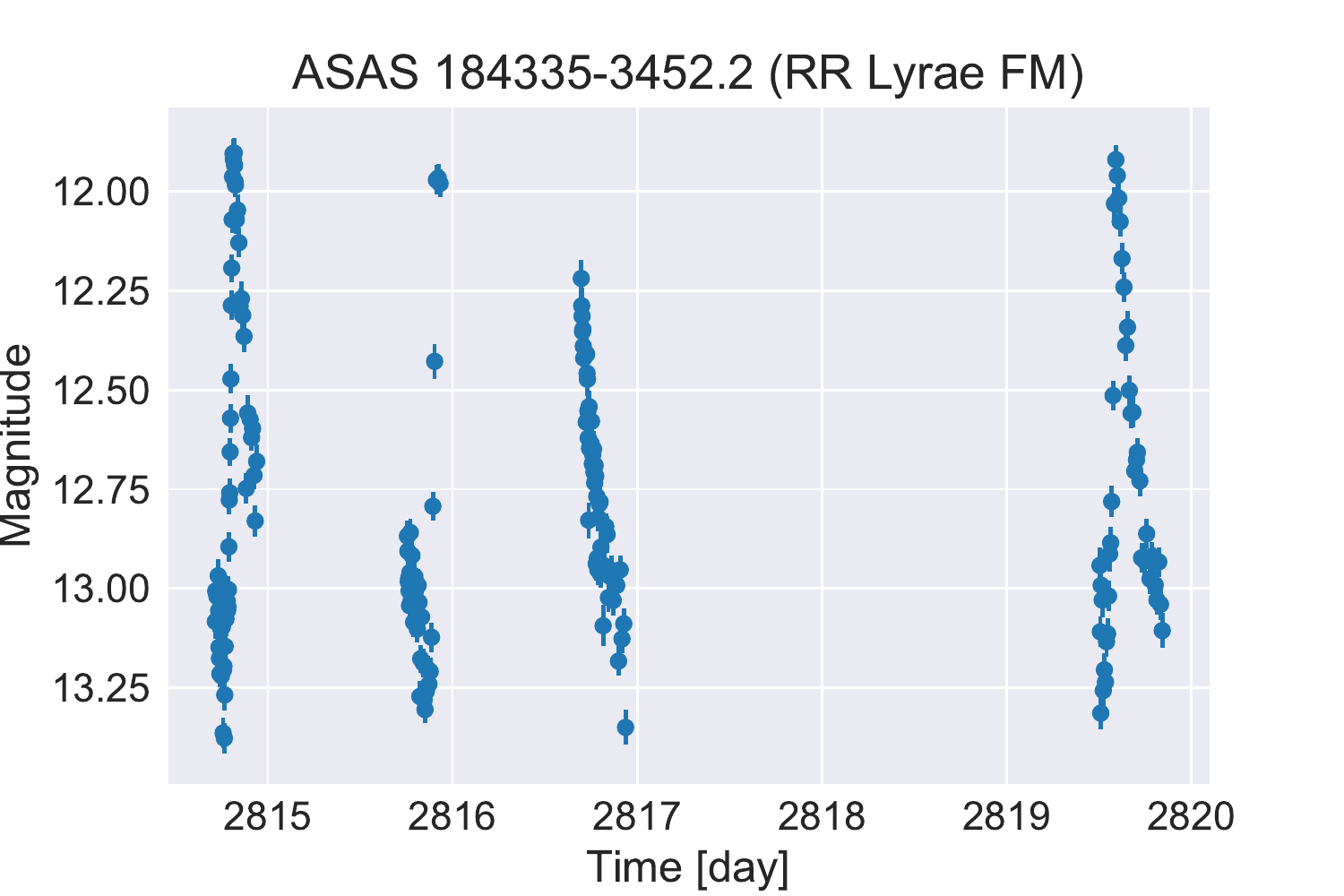}
        \caption{25th percentile}
    \end{subfigure}
    \begin{subfigure}[b]{0.495\textwidth}
        \includegraphics[width=\textwidth]{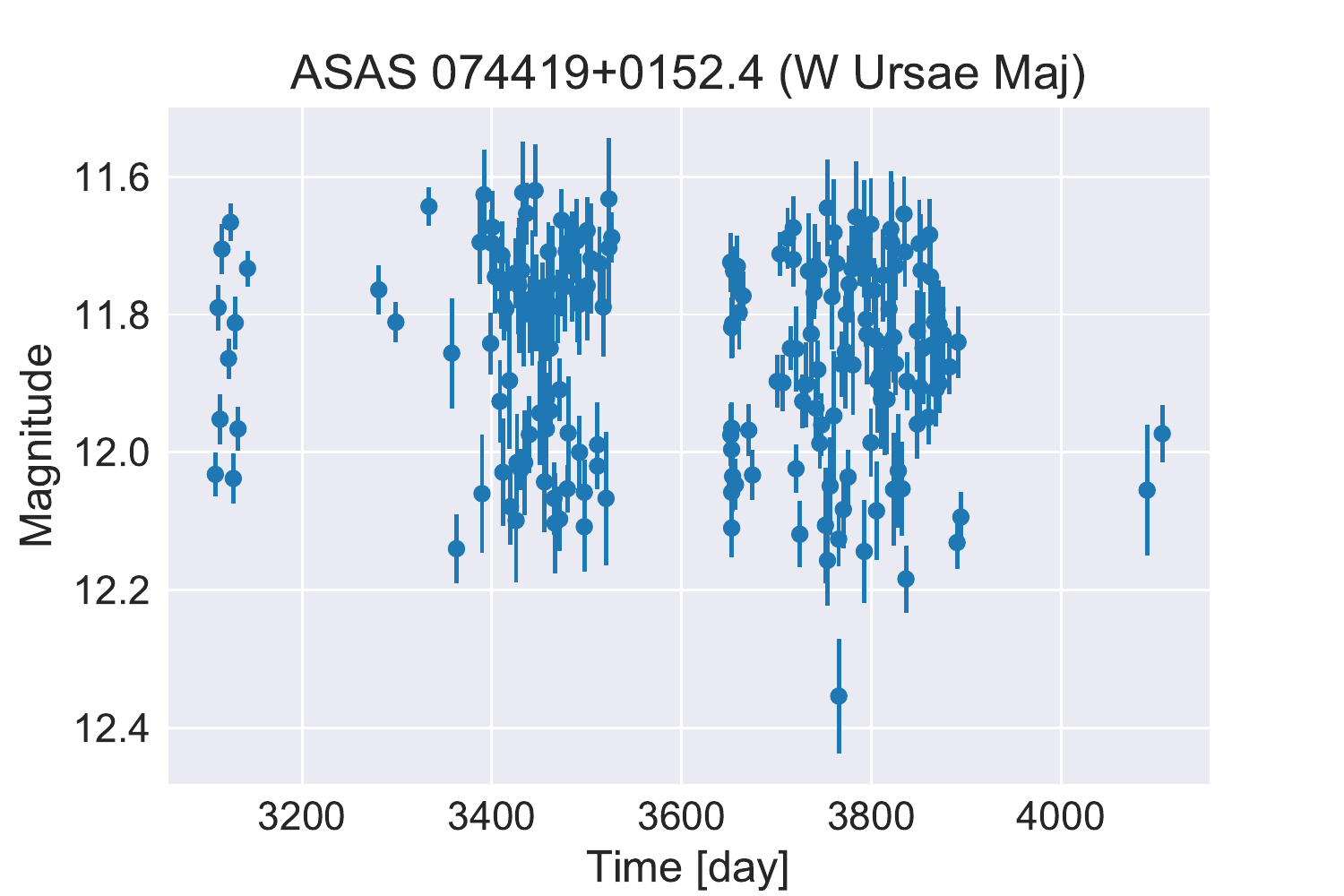}
        \caption{75th percentile}
    \end{subfigure}\\
    \begin{subfigure}[b]{0.495\textwidth}
        \includegraphics[width=\textwidth]{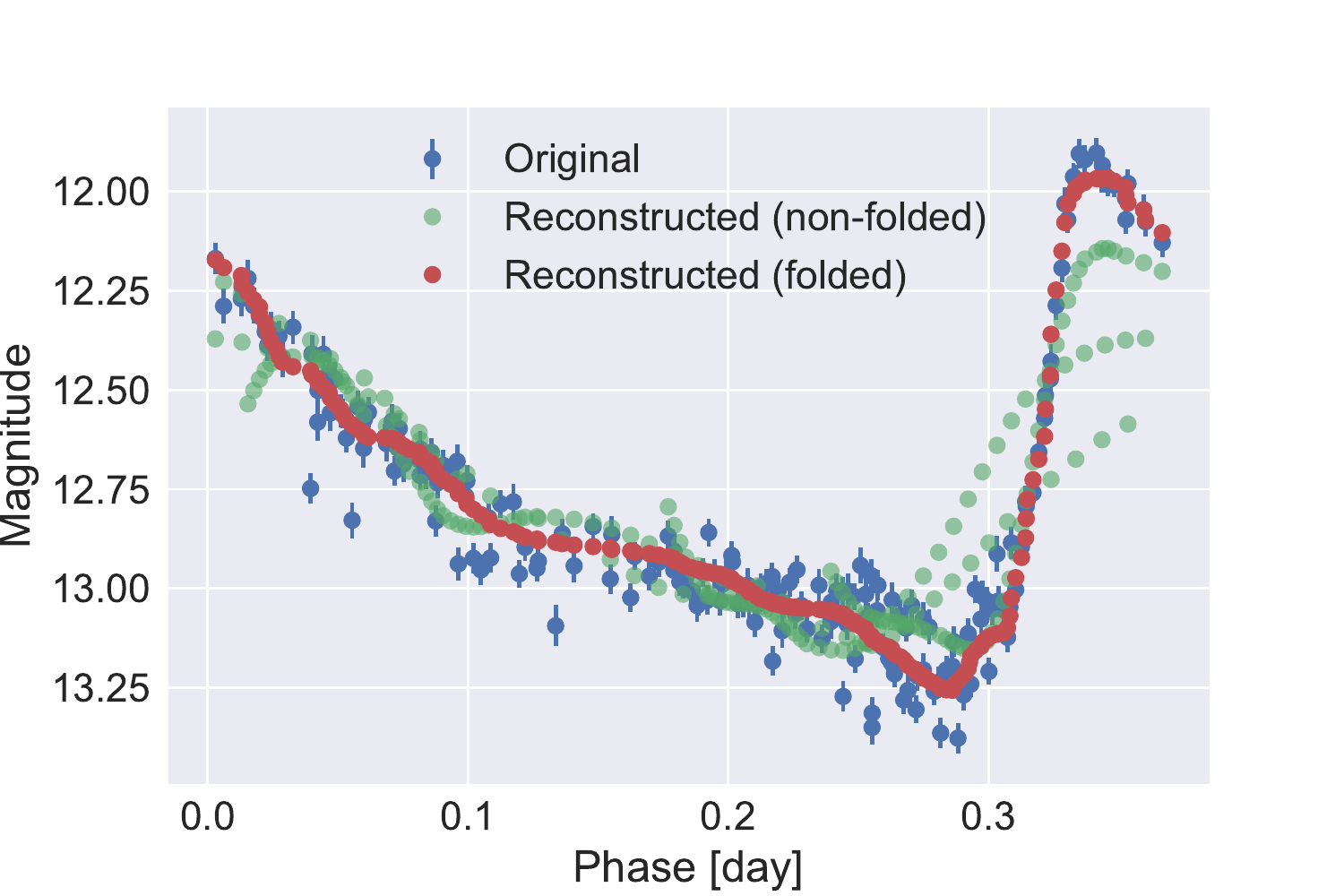}
        \caption{25th percentile (folded)}
    \end{subfigure}
    \begin{subfigure}[b]{0.495\textwidth}
        \includegraphics[width=\textwidth]{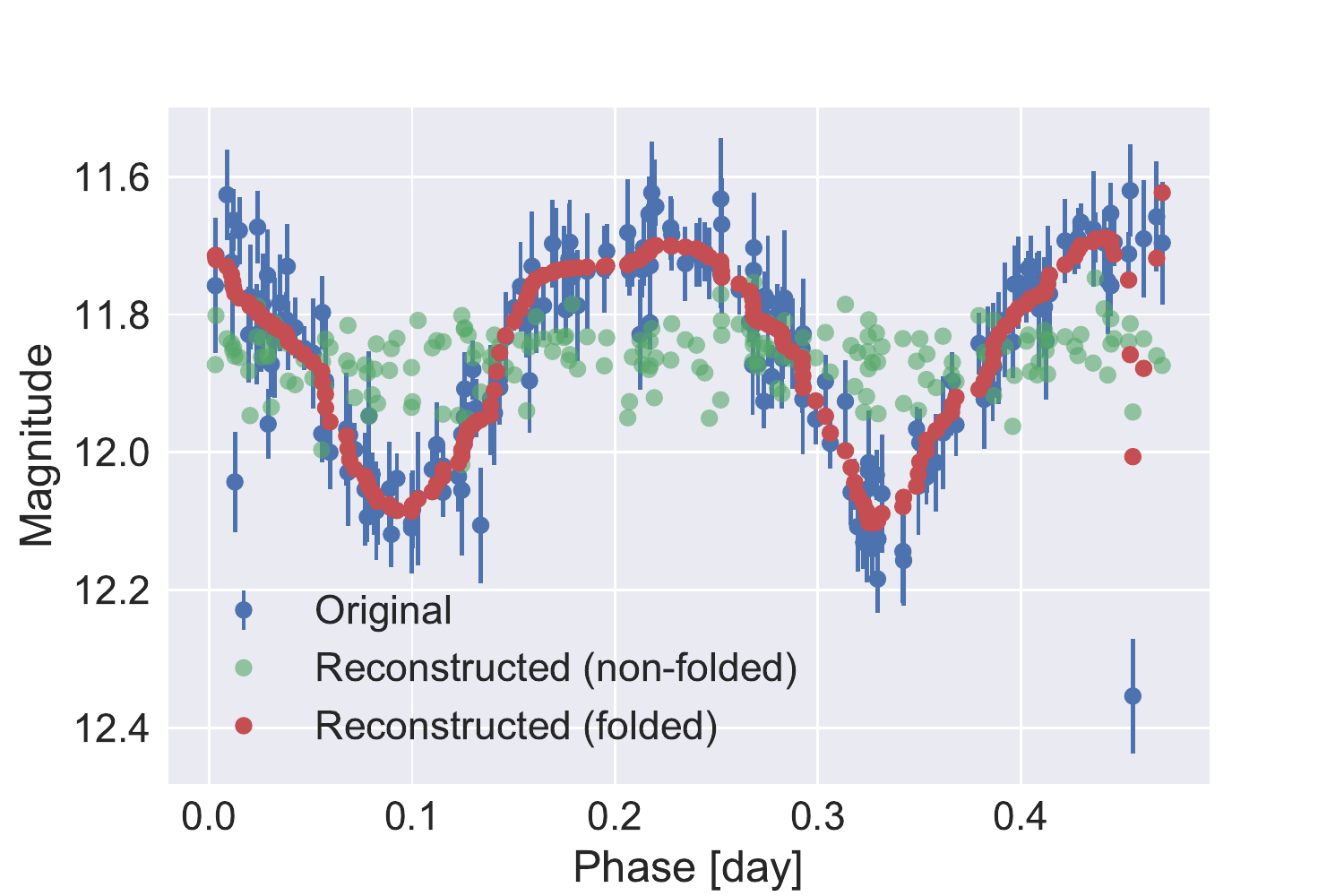}
        \caption{75th percentile (folded)}
    \end{subfigure}
    \caption{\footnotesize
        \textbf{Example autoencoder reconstructions of ASAS light curves from
        64-dimensional feature representation.} Twenty-fifth and seventy-fifth percentile
        error reconstructions are shown for raw, unfolded light curves in a) and b), and
        for period-folded light curves in c) and d).
        The $\sim$V-band magnitudes are in the Vega system.}
    \label{fig:asas_reconstruct}
\end{figure}

\begin{figure}[htb!]
    \centering
    \begin{subfigure}[b]{0.45\textwidth}
        \includegraphics[width=\textwidth]{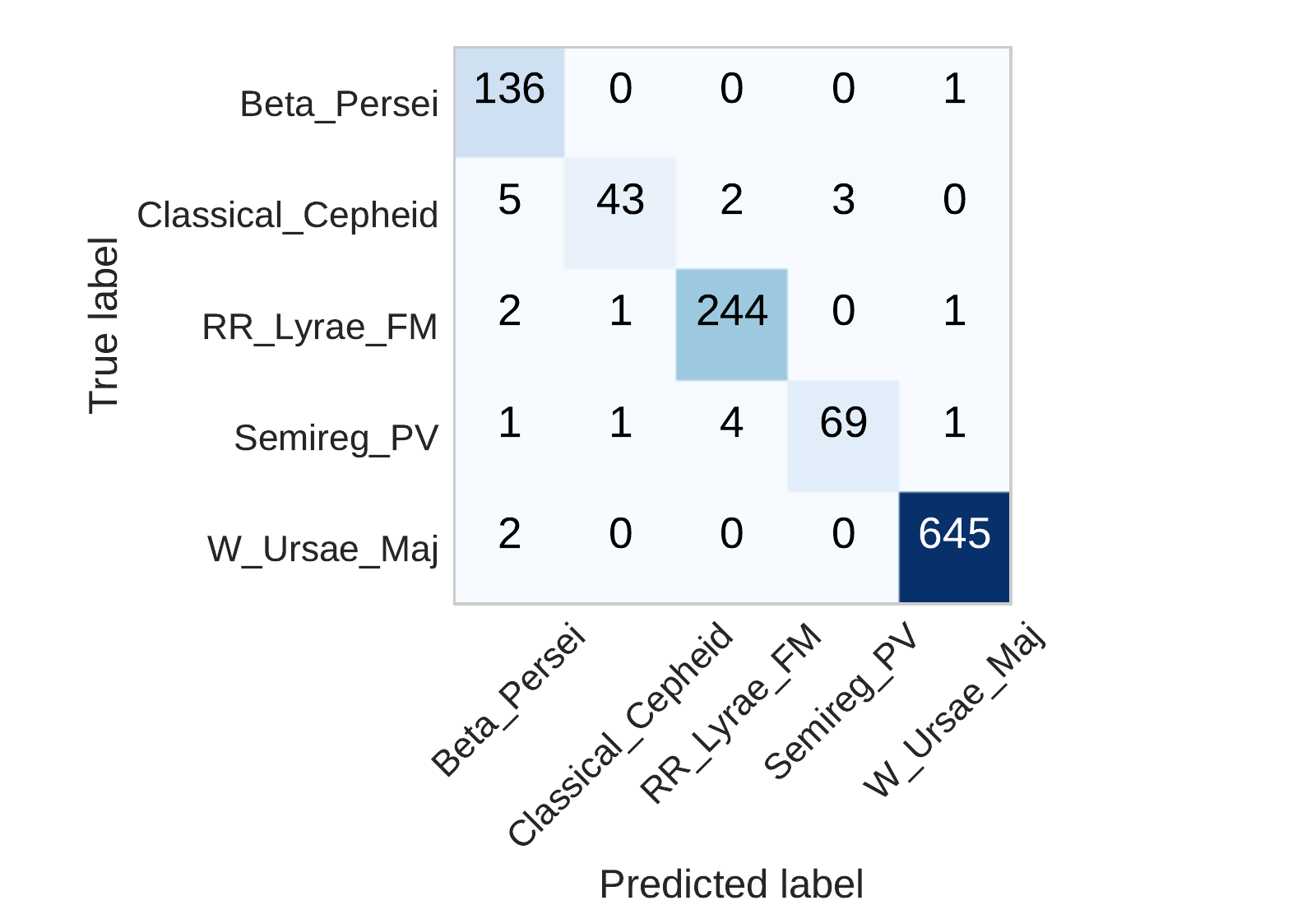}
        \caption{ASAS}
        \label{fig:asas_confusion}
    \end{subfigure}
    \begin{subfigure}[b]{0.45\textwidth}
        \includegraphics[width=\textwidth]{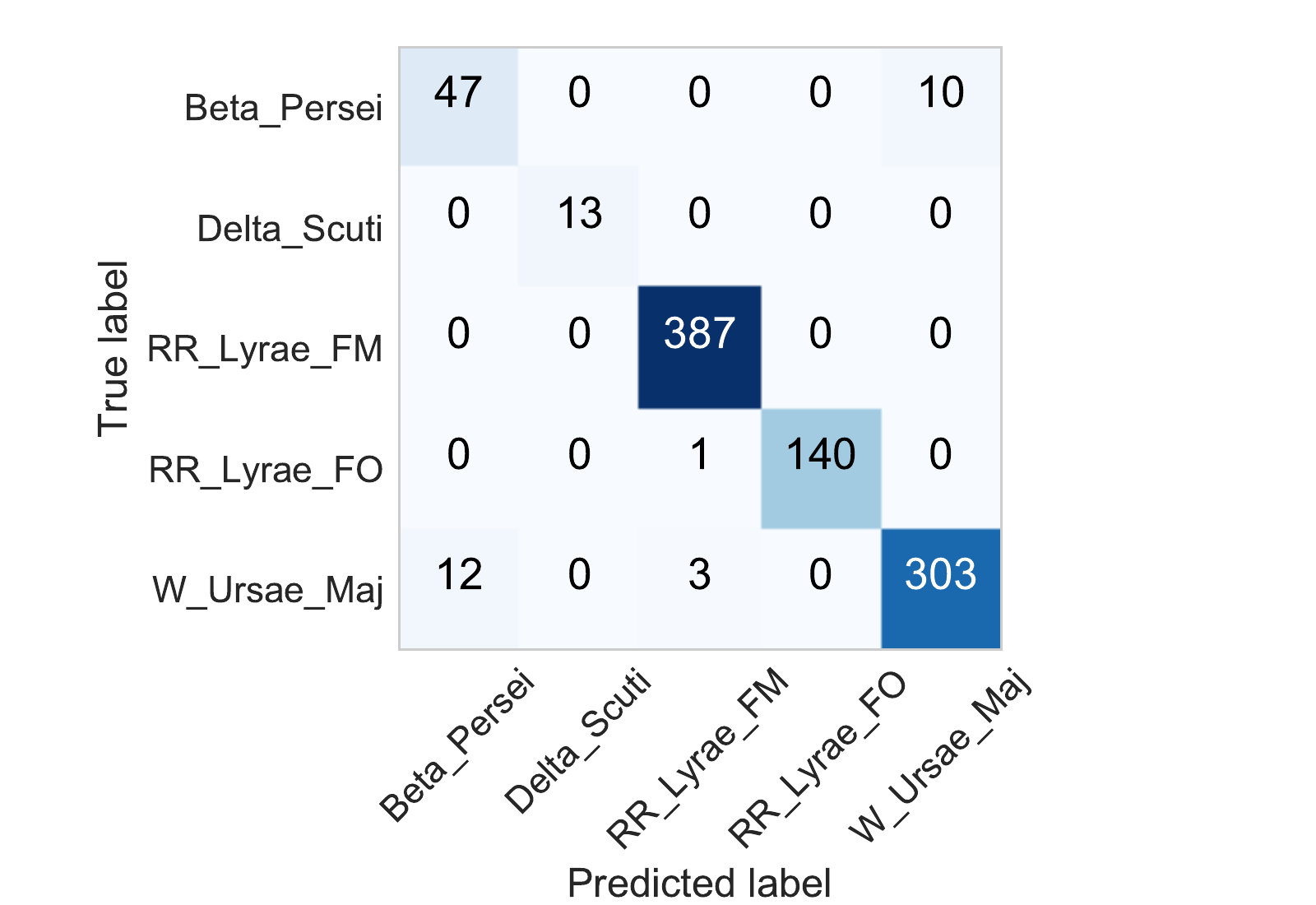}
        \caption{LINEAR}
        \label{fig:linear_confusion}
    \end{subfigure}
    \begin{subfigure}[b]{0.45\textwidth}
        \includegraphics[width=\textwidth]{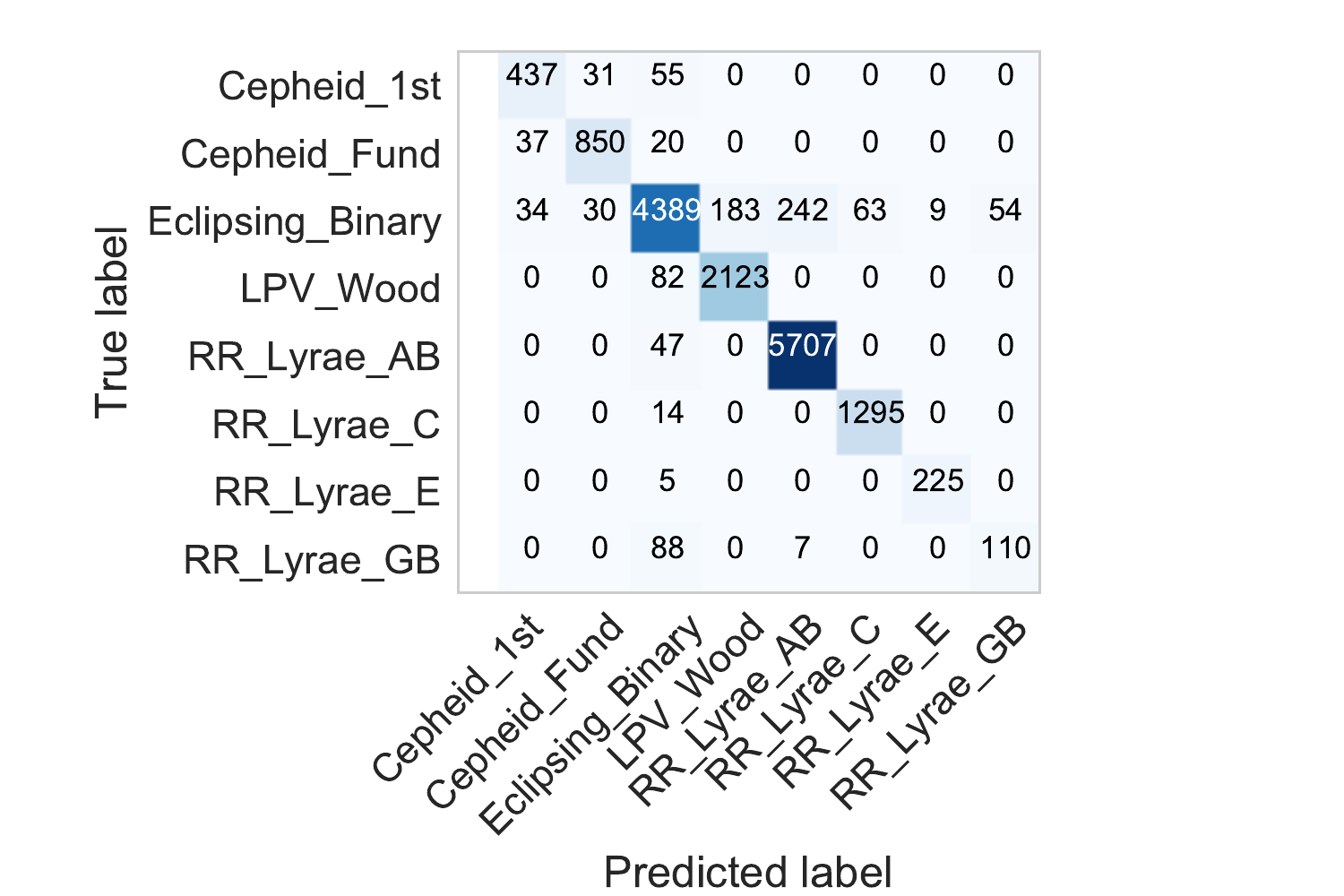}
        \caption{MACHO}
        \label{fig:macho_confusion}
    \end{subfigure}
    \caption{\textbf{Confusion matrices for autoencoder-feature random forest
    classifiers for labeled variable star light curves from a) ASAS, b) LINEAR, and c)
    MACHO surveys.} Values along the diagonals are counts of correctly-classified light
    curves, and off-diagonal values correspond to incorrect classifications (darker squares
    correspond to higher counts).}
    \label{fig:confusion}
\end{figure}

\begin{table}[h!]
\centering
 \begin{tabular}{|c|ccc|} 
 \hline
 \backslashbox{Method}{Dataset} & ASAS & LINEAR & MACHO \\
 \hline
 Autoencoder & $98.77\%\pm0.39\%$ & \bm{$97.10\%\pm0.60\%$} & \bm{$93.59\%\pm0.15\%$} \\
 Richards et al. & \bm{$99.42\%\pm0.27\%$} & $96.72\%\pm0.67\%$ & $90.50\%\pm0.22\%$ \\
                 & ($+0.66\pm0.49$)\% 
                 & ($-0.37\pm0.45$)\% 
                 & ($-2.74\pm0.16$)\% \\                   
 Kim/Bailer-Jones & $98.83\%\pm0.33\%$ & $95.49\%\pm0.83\%$ & $88.98\%\pm0.19\%$ \\
  &  ($+0.06\pm0.32$)\% 
  &  ($-1.60\pm0.59$)\% 
  &  ($-4.60\pm0.16$)\% \\
 \hline
 \end{tabular}
 \caption{\textbf{Validation accuracies (mean $\pm$ standard deviation across five
stratified cross-validation splits) for the autoencoder, Richards et al., and
Kim/Bailer-Jones feature random forests.} Note that the training labels for the ASAS data 
were identified in part using the Richards et al.\ features. In parentheses are the mean and standard deviation of the differences in accuracy (Other Method - Autoencoder; a negative value means the autoencoder performed better) over the cross-validation folds.}
 \label{tab:accuracy}
\end{table}

\section*{Methods}
We describe here the implementation specifics of the proposed neural network
classifier, and the detailed properties of the datasets which were used in the above
experiments.

\subsection*{Comparison with other neural network approaches}
Many of the most commonly used approaches for time series analysis, including standard
recurrent neural networks, rely on an implicit assumption that the data is uniformly
sampled in time.
This assumption is not unique to neural network approaches: the Fast
Fourier Transform (FFT), commonly used in featurization, is only well-defined for the
evenly sampled, homoskedastic case.
Existing neural networks that do allow for uneven sampling operate on interpolations of
the data (see, e.g.,~\cite{charnock2016deep,che2016recurrent,lipton2015learning}), thereby
replacing the problem with one that can be solved by standard approaches. Any form of
interpolation makes implicit assumptions about the spectral structure of the data, which
may be unjustified and can end up introducing biases and artifacts. These artifacts
increase with sampling unevenness and are ultimately properties of the algorithm, not the
data.

\subsection*{Loss function for known measurement errors}
Typically an autoencoder is trained to minimize the difference between the input
and the reconstructed values, usually in terms of mean squared error.
In the case of astronomical surveys, individual measurement errors are generally
available for each time step. We therefore constructed a new loss function which weighs
the reconstruction error at each time step more or less heavily when the measurement errors are
small or large, respectively
(in analogy with standard weighted least squares regression). In particular, our models
are trained to minimize the weighted mean squared error 
\begin{equation}
    \textrm{WMSE} = \frac{1}{n_T} \sum_{i=1}^{N} \sum_{j=1}^{n_T}
    \left(y_i^{(j)} - \hat{y}_i^{(j)}\right)^2/\left(\sigma_i^{(j)}\right)^2,
    \label{eq:wmse}
\end{equation}
where $n_T$ is the length of the sequences, $N$ the number of light curves, and $y_i^{(j)},$
$\hat{y}_i^{(j)},$ and $\sigma_i^{j}$ are (respectively) the $j$th measurement,
reconstruction value, and measurement error of the $i$th light curve.

\subsection*{Neural network parameters}
The autoencoders used for the ASAS and LINEAR survey classification tasks were constructed
using a network architecture like that of Fig.~\ref{fig:network_auto}, consisting of two
encoding and two decoding GRU layers of size 96, and an embedding size of 64. The network
is trained using the Adam optimizer with learning rate $\lambda=5\times10^{-4}$ to
minimize the weighted mean squared reconstruction error defined in Eq.~(\ref{eq:wmse}).

\subsection*{Random forest classifier parameters}
In the classification experiments, a random forest classifier is trained to predict the class
of each labeled light curve from either the unsupervised autoencoder features from our
method, or the baseline features from~\cite{richards2011machine}.
The hyperparameters of the random forest were chosen by performing a five-fold cross
validation grid search over the following grid: $n_{\textrm{trees}} \in \{50,100,250\},
\textrm{criterion} \in \{\textrm{gini,entropy}\}, \textrm{max features} \in \{3,6,12,18\},
\textrm{min leaf samples} \in \{1,2,3\}$. In each case, 80\% of the available samples are
used as training data, and 20\% are withheld as test data (the hyperparameter selection
is performed only using the training data); the same splits are used for the autoencoder
and Richards et al.\ features, and in each case we present the results across five
different choices of train/test splits.

\subsection*{All Sky Automated Survey (ASAS) data}
The ASAS Catalog of Variable stars~\cite{pojmanski2002all} consists of 50,124
(mostly unlabeled) variable star light curves. For the ASAS dataset, we select 349 Beta
Persei, 130 Classical Cepheids, 798 RR Lyrae (FM), 184 Semiregular PV, and 181 W Ursae
Major class stars.
The class label of each star is either either manually identified or predicted with high
probability ($>$99\%) by the Machine-learned ASAS Classification Catalog, and periods used
in period-folding were identified programmatically as described
in~\cite{richards2012construction}.

\subsection*{Lincoln Near-Earth Asteroid Research (LINEAR) data}
The LINEAR dataset consists of 5,204 light curves from five classes of star: 2,234
RR Lyrae FM, 1,860 W Ursae Major, 749 RR Lyrae FO, 291 Beta Persei, and 70 Delta Scuti.
All of the light curves in the LINEAR dataset were manually
classified, and periods used for period-folding were validated manually
as described in~\cite{palaversa2013exploring}.

\subsection*{Massive Compact Halo Object (MACHO) Project data}
The MACHO dataset consists of 21,474 light curves from eight classes of star: 7,405 RR
Lyrae AB, 6,835 Eclipsing Binary, 3,049 Long-Period Variable Wood (subclasses A-D were
combined into a single superclass), 1,765 RR Lyrae C, 1,185 Cepheid Fundemantal, 683
Cepheid First Overtone, 315 RR Lyrae E, and 237 RR Lyrae/GB Blend.
All models were trained on brightness and error values from the red band, though the same
approaches could be used on the blue band or both bands simultaneously.
Periods and labels were determined using a semi-automated procedure as described
in~\cite{alcock1996macho}.
The full MACHO dataset also contains light curves from many more non-variable sources, which
were not used in training either the autoencoder or the random forests in our experiments.

\subsection*{Data preprocessing}
We first processed the raw data from each survey as described in
\cite{richards2012construction}, removing low-quality measurements (those with grade C or
below), so that each source consists of a series of observations of time, brightness
(V-band magnitude), and measurement error values, denoted $(\bm t_i, \bm y_i, \bm
\sigma_i)$. Before training, we further preprocessed the data by centering and scaling
each light curve to have mean zero and standard deviation one. We also manually removed
light curves from our autoencoder training set which did not exhibit any notable periodic
behavior by computing a ``super smoother''~\cite{friedman1989flexible} fit for each light
curve with period equal to the estimated period from~\cite{richards2012construction} and
omitting light curves with residual greater than 0.7 (we observed that including aperiodic
sources tended to degrade the quality of the reconstructions). Finally, we partitioned the
light curves into subsequences of length 200; this is not strictly necessary since our
recurrent architecture allows for input and output sequences of arbitrary length, but the
use of sequences of equal length is computationally advantageous and reduces training time
(see SI for details).
The resulting dataset consists of 33,103 total sequences of length $n_T=200$.

\section*{Data availability statement}
All data and code for reproducing the above experiments is available online at
\url{https://github.com/bnaul/IrregularTimeSeriesAutoencoderPaper}, including Python code
implementing the simulations, Jupyter notebooks for reproducing the figures, and trained
autoencoder models and weights.

\centerline{\large{\bf{Supplementary information (SI)}}}

\renewcommand{\figurename}{Supplementary Figure}
\renewcommand{\tablename}{Supplementary Table}

\section{Background on neural networks for time series data}

Machine learning techniques for time series data typically map each sequence to a set of
scalar-valued features that attempt to capture various distinguishing properties; these
may range from simple summary statistics like median or maximum to parameters of
complicated model fits. Features are then used as inputs to train models which map
fixed-length feature vectors to the relevant labels. The process of building and
selecting features can be difficult, requiring extensive expert knowledge and iterative
fine-tuning, and feature generation itself may be computationally expensive.
One distinct advantage of neural networks is the elimination of the need to manually
create, compute, and select effective features~\cite{lecun2015deep}.
Nearly all applications of neural networks to time series data rely on the assumption that
samples are evenly spaced in time.

Artificial neural networks have previously been applied to many time series tasks, including
classification and forecasting.
The simplest approaches make use of fully connected networks
(see, e.g.,~\cite{lapedes1987nonlinear}), which treats each sequence as consisting of
independent observations and thereby ignores the local temporal structure.
Convolutional neural networks have also been applied to time series data with considerable
success~\cite{lecun1995convolutional}, in particular for the problem of speech recognition
\cite{hinton2012deep}.
The enormous popularity of convolutional networks for
image analysis has led to efficient optimization techniques and highly optimized software
implementations, so convolutional networks tend to be faster and easier to train than
other approaches, including recurrent networks (which our technique employs).
As opposed to fully connected networks which ignore the local structure of sequences,
convolutional networks are well-suited for recognizing local, short-term patterns in
time series data. However, convolutional networks are less ideally suited for handling
irregular sequences of unequal length.
For this reason we focus on recurrent networks, which are designed to
handle sequences of arbitrary length.

Recurrent networks have been found to be more difficult to train than convolutional
networks due to their repeating structure, which makes them more susceptible to problems
of exploding or vanishing gradients during training~\cite{pascanu2013difficulty}.
Another drawback of the recursive structure of recurrent networks is that recomputing the
internal state for each input can lead to rapid loss of information over time and
difficulty in tracking long-term dependencies; these difficulties led to the development
of the popular long short-term memory (LSTM)~\cite{hochreiter1997long} and gated recurrent
unit (GRU)~\cite{cho2014learning} architectures, which help information be retained over
longer periods of time.
Some sequence processing tasks for which recurrent neural networks have been
successfully applied include recognition~\cite{graves2013speech}, machine translation
\cite{bahdanau2014neural}, and natural language processing~\cite{mikolov2010recurrent}.

\section{Frequency-domain characteristics of unevenly sampled time series}\label{sec:freq}
One important class of features that is often used in time series inference problems is
frequency-domain properties, including estimates of period(s) or power spectra over some
range of frequencies. By far the most common approach for transforming time series data
between the time and frequency domains is the (discrete) Fourier transform (DFT).
The default formulation of the DFT assumes that the data are uniformly sampled. If this
assumption isn't met, more complex approaches are required that interpolate the data onto
a uniform grid~\cite{beylkin1995fast} or project it onto spaces with fixed, predefined
bandwidth. A frequently used version of the latter approach is the Lomb-Scargle
periodogram~\cite{lomb1976least,scargle1982studies}, which estimates the frequency
properties of a signal using a least squares fit of sinusoids. Once the power spectrum of
a signal has been estimated using one of the above methods, features such as the dominant
frequencies/periods can be extracted and used in machine learning tasks such as
classification.

In the next section, we demonstrate a neural network architecture which is able to infer
periodic behavior directly from an irregularly sampled time series without constructing an
explicit estimate of the full power spectrum.
\section{Simulation study: sinusoidal data}\label{sec:simulation}
In order to evaluate the effect of depth, hidden layer size, and other architecture
choices, we first carry out a number of experiments using simulated time series data
generated from a know distribution. In particular, we construct periodic functions of the
form
\begin{equation}
f_i(t) := A_i \sin(2 \pi \omega_i t + \phi_i) + b_i
\label{eq:sinusoid}
\end{equation}
for randomly selected parameter values $\omega_i$ (frequency; we will also write
$T_i=\omega_i^{-1}$ to denote period), $A_i$ (amplitude), $\phi_i$
(phase), and $b_i$ (offset). The parameter values are chosen as independent
identically-distributed (i.i.d.) samples from the probability distributions
\begin{align}
    A_i &\sim \mathcal{U}(0.5, 2), \label{eq:sinusoid_A}\\
    \omega_i^{-1} = T_i &\sim \mathcal{U}(1, 10), \label{eq:sinusoid_omega}\\
    \phi_i &\sim \mathcal{U}(-\pi, \pi), \label{eq:sinusoid_phi}\\
    b_i &\sim \mathcal{N}(0, 1),\label{eq:sinusoid_b}
\end{align}
where $\mathcal{U}(a, b)$ is the uniform distribution on $[a,b]$, and $\mathcal{N}(\mu,
\sigma^2)$ is the normal distribution with mean $\mu$ and variance $\sigma^2$.
For each periodic function, we generate a set of $n_T=200$ sampling times
$(t_i^{(1)}, \dots, t_i^{(n_T)})$ by sampling the differences from a heavy-tailed power
law distribution $t_i^{(j+i)} - t_i^{(j)} \sim \operatorname{Lomax}(0.05, 2)$, where the
$\operatorname{Lomax}(\lambda, \alpha)$ distribution has probability density function
\begin{equation*}
    f(x; \lambda, \alpha) =
    \frac{\alpha}{\lambda}\left(1 + \frac{x}{\lambda}\right)^{-\left(\alpha+1\right)}.
\end{equation*}
The resulting distribution of times is such that the mean total time is
$E[t_{n_T}]=10$ but the sequence $(t_i^{(1)}, \dots, t_i^{(n_T)})$ is highly unevenly
sampled and may contain many large gaps between consecutive samples.

\begin{figure}[htpb]
    \centering
    \begin{subfigure}[b]{0.32\textwidth}
        \includegraphics[width=\textwidth]{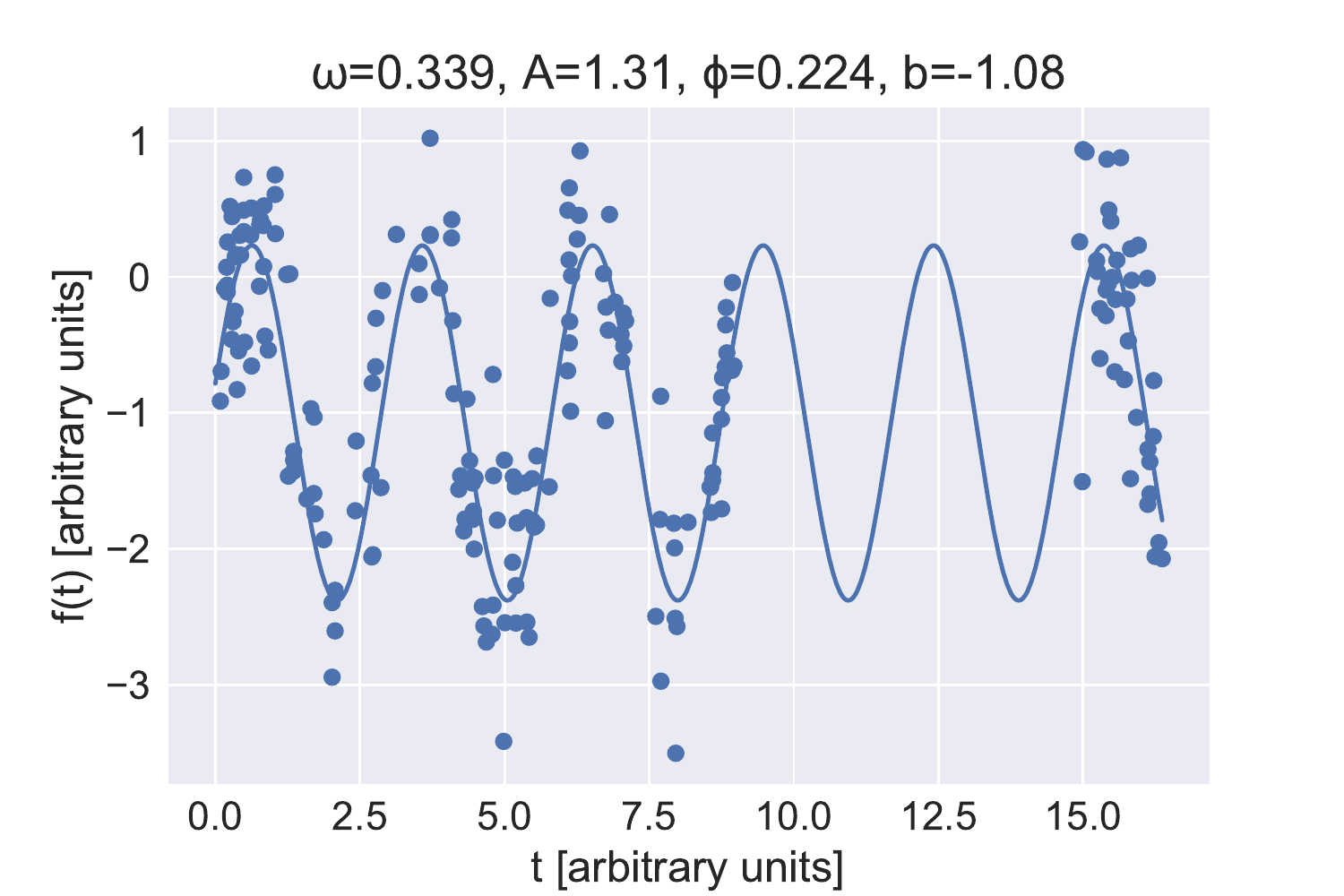}
        \caption{}
    \end{subfigure}
    \begin{subfigure}[b]{0.32\textwidth}
        \includegraphics[width=\textwidth]{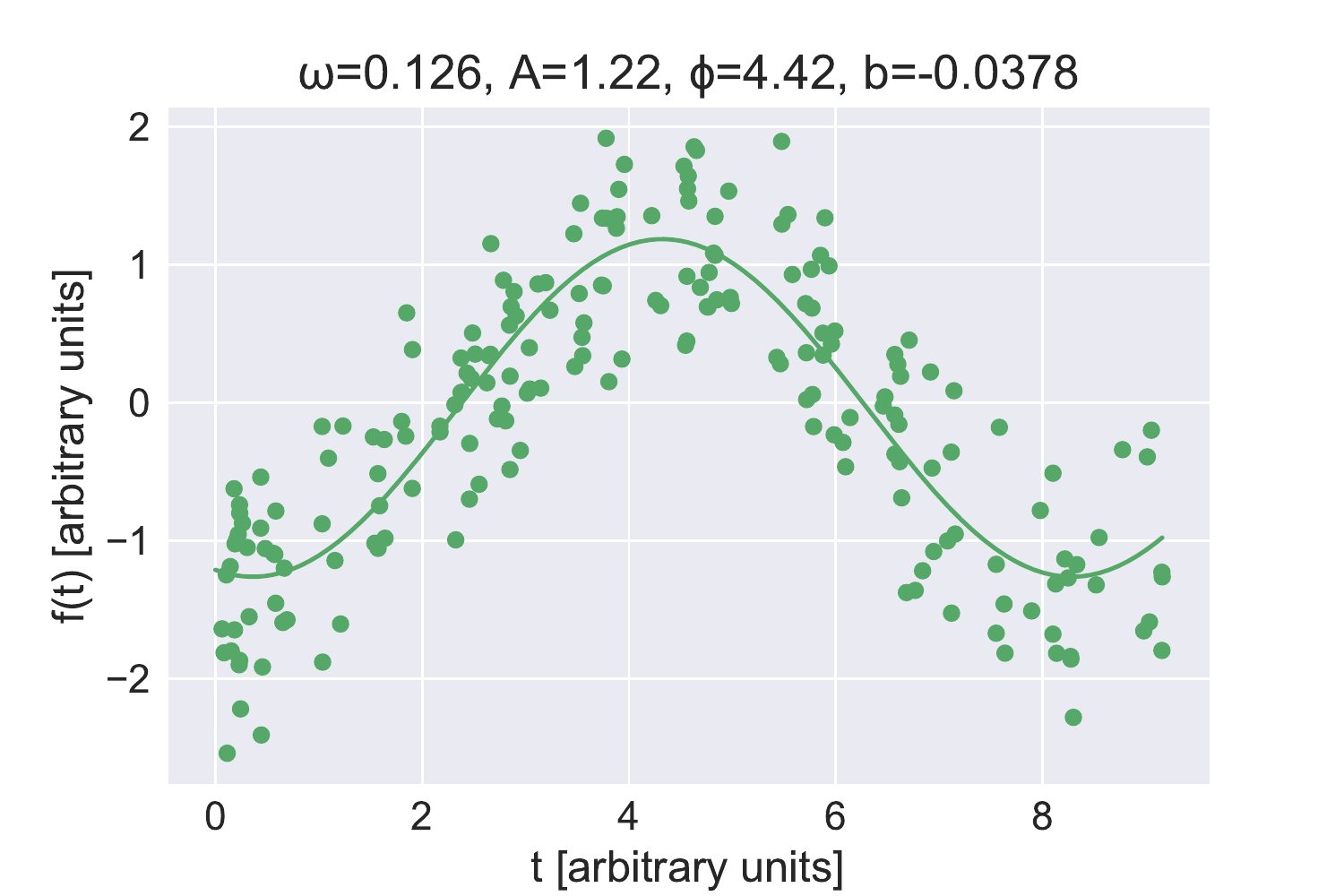}
        \caption{}
    \end{subfigure}
    \begin{subfigure}[b]{0.32\textwidth}
        \includegraphics[width=\textwidth]{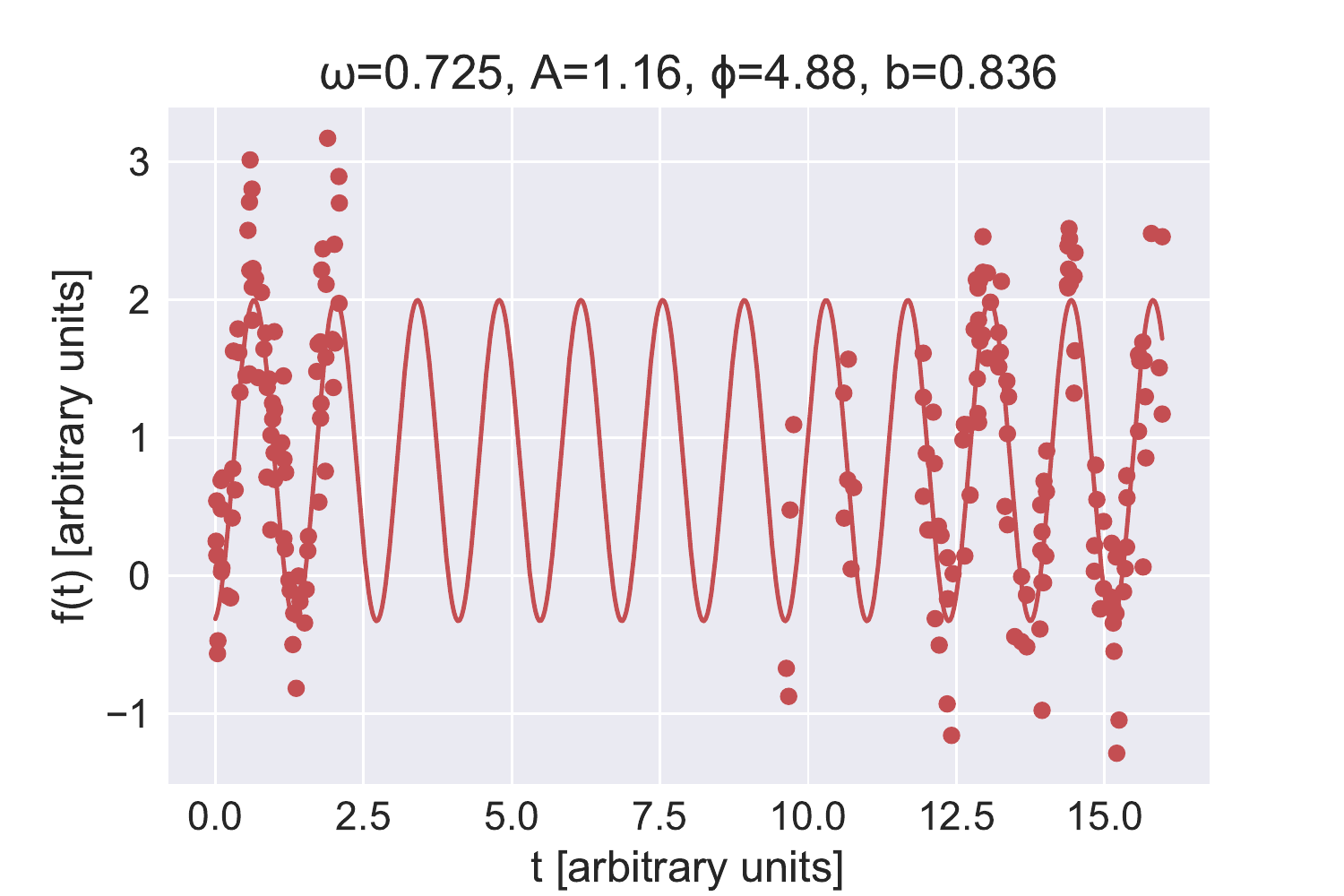}
        \caption{}
    \end{subfigure}
    \caption{\textbf{Examples of randomly generated periodic functions.} Panels a), b) and
c) each contain values of a sinusoidal function with random parameters evaluated at
random times. The distributions of the parameters are given in
\eqref{eq:sinusoid_A}--\eqref{eq:sinusoid_time}.}
    \label{fig:sinusoid}
\end{figure}

Our training dataset consists of $N_{\textrm{train}} = 50,000$ periodic functions and
sampling time sequences generated using the above procedures; we also add white Gaussian random
noise $\bm \epsilon_i^{(j)} \sim \mathcal{N}(0, \sigma^2)$ with $\sigma=0.5$ to each
measurement, so the final input data has the form
\begin{equation}
    \bm t_i = (t_i^{(1)}, \dots, t_i^{(n_T)}),
    \ \bm y_i = (f_i(t_i^{(1)}) + \epsilon_i^{(1)}, \dots, f_i(t_i^{(n_T)}) + \epsilon_i^{(1)}).
    \label{eq:sinusoid_time}
\end{equation}
Supplementary Figure~\ref{fig:sinusoid} shows examples of randomly generated periodic functions and
randomly selected points with added noise.

The accuracies and losses for each model are evaluated using a separate dataset of
$N_{\textrm{valid}} = 10,000$ sequences generated using the same methodology.
\subsection{Estimating period, phase, amplitude, offset}
As our initial experiment, we implement a network which solves a supervised regression
problem of estimating the parameters $\bm \theta_i = (\omega_i, A_i, \phi_i, b_i)$ [as defined
in~\eqref{eq:sinusoid}] from the generated times and measurement values. The corresponding network
consists of the encoder portion of the network shown in Supplementary Figure~1 of the main text
with an output size of four. The network is trained to minimize the mean squared error
\begin{equation}
\textrm{MSE} = \frac{1}{N_{\textrm{train}}} \sum_{i=1}^{N_{\textrm{train}}} \|\theta_i -
\hat{\theta}_i\|_2^2
\end{equation}
for the given training data; the target values are first preprocessed by re-parametrizing
$\bm \theta_i$ as $(T_i, A_i \cos \phi_i, A_i \sin \phi_i, b_i)$ (which leads to
faster convergence) and then centering and scaling each target variable to have mean 0 and
standard deviation 1. Our network is trained via the Adam optimization method
\cite{kingma6980method} with standard parameter values $\beta_1=0.9, \beta_2=0.999$ using a
learning rate of $\eta=5\times10^{-4}$ and a batch size of 500. We also impose 25\% dropout
\cite{srivastava2014dropout} between recurrent layers for regularization, which does not
seem to be necessary for this simple task but is beneficial for the astronomical classification tasks discussed in the main text. All models were implemented in Python using the Keras library
\cite{chollet2015keras} and trained using an NVIDIA Tesla K80 GPU.

\begin{figure}[htpb]
\centering
\begin{subfigure}[b]{0.495\textwidth}
\includegraphics[width=\textwidth]{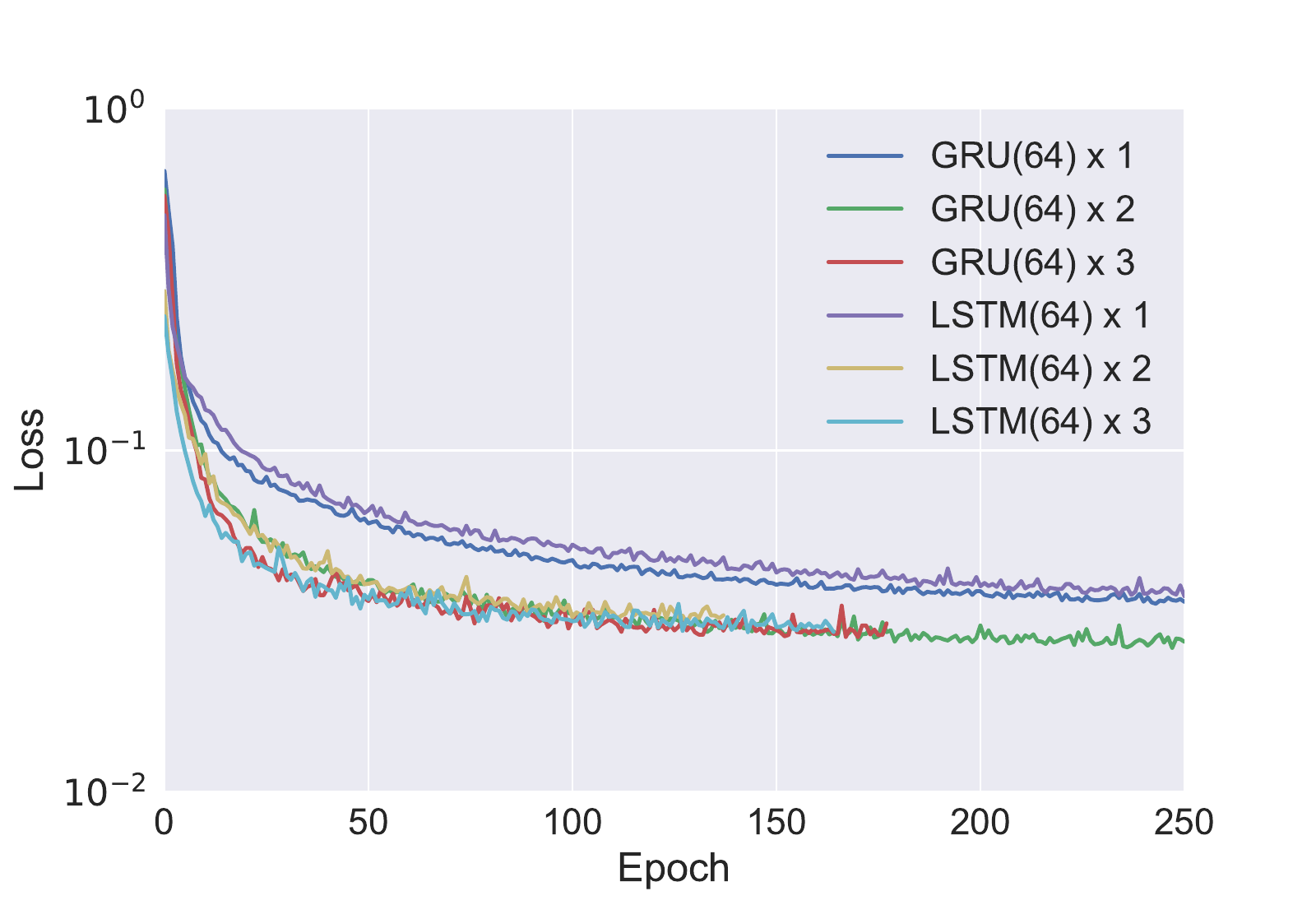}
\caption{By epoch}
\end{subfigure}
\begin{subfigure}[b]{0.495\textwidth}
\includegraphics[width=\textwidth]{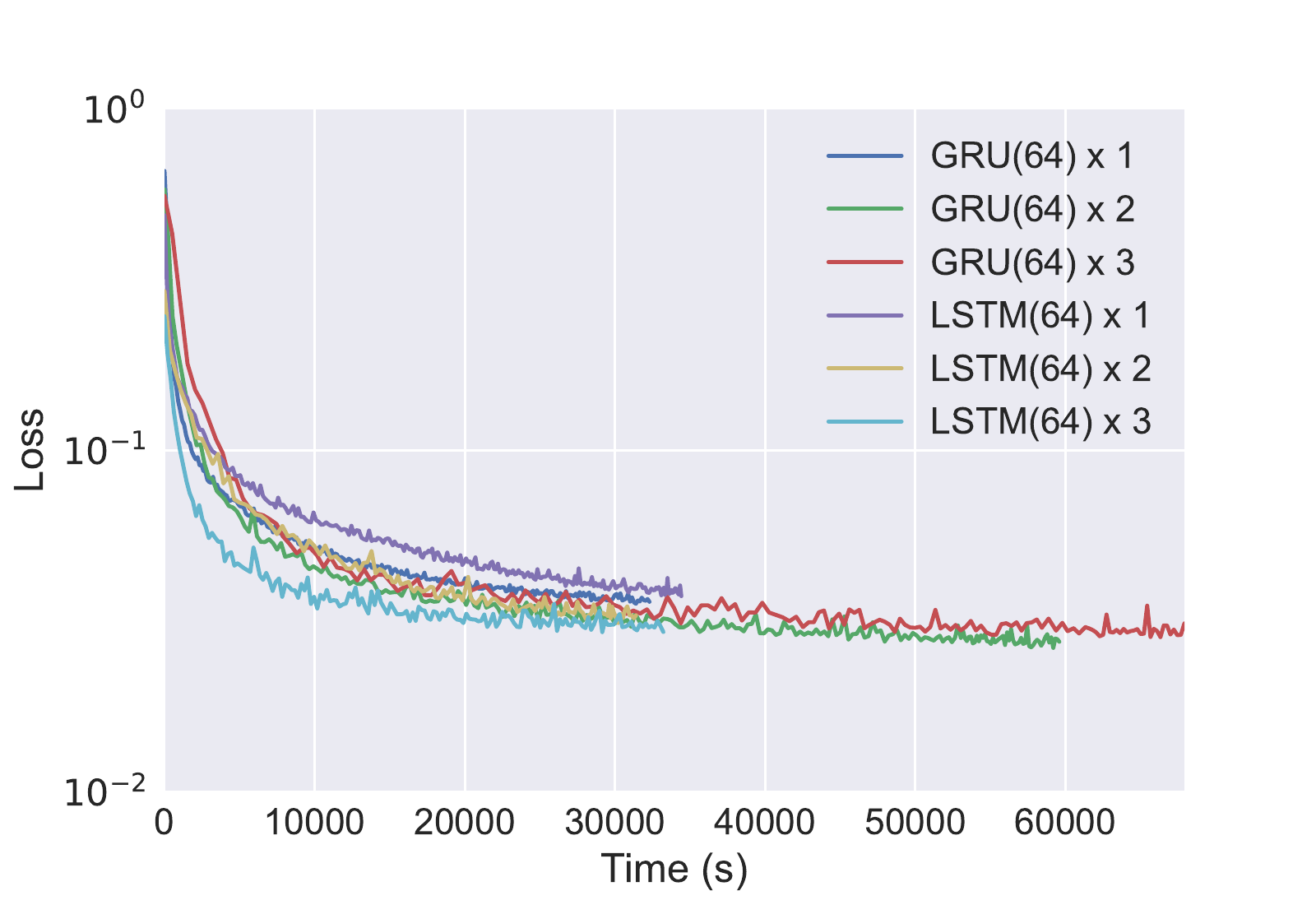}
\caption{By time}
\end{subfigure}
\caption{\textbf{Comparison of validation losses during training for various numbers and
types of layers on the periodic parameters $\bm \theta_i$.} The loss plotted is the mean
squared error over the validation set of the parameters being estimated as a function of
a) training epoch and b) training wall time.  Each colored line corresponds to a specific
network architecture (layer type, layer size, and network depth) as shown in the legend:
for example, ``GRU(64) $\times$ 2'' denotes two decoder GRU layers each composed of 64
hidden units.}
\label{fig:gru_lstm_period}
\end{figure}

As seen in Supplementary Figures~\ref{fig:gru_lstm_period} and~\ref{fig:gru_period}, the maximum
accuracy achieved by the one layer networks is somewhat inferior to
that of the multi-layer networks; nevertheless, it is clear that all of the architectures
considered are able to accurately recover information about the periodic characteristics
of the unknown functions, even in the presence of noise and sampling irregularity.
\begin{figure}[htpb!]
\centering
\begin{subfigure}[b]{0.495\textwidth}
\includegraphics[width=\textwidth]{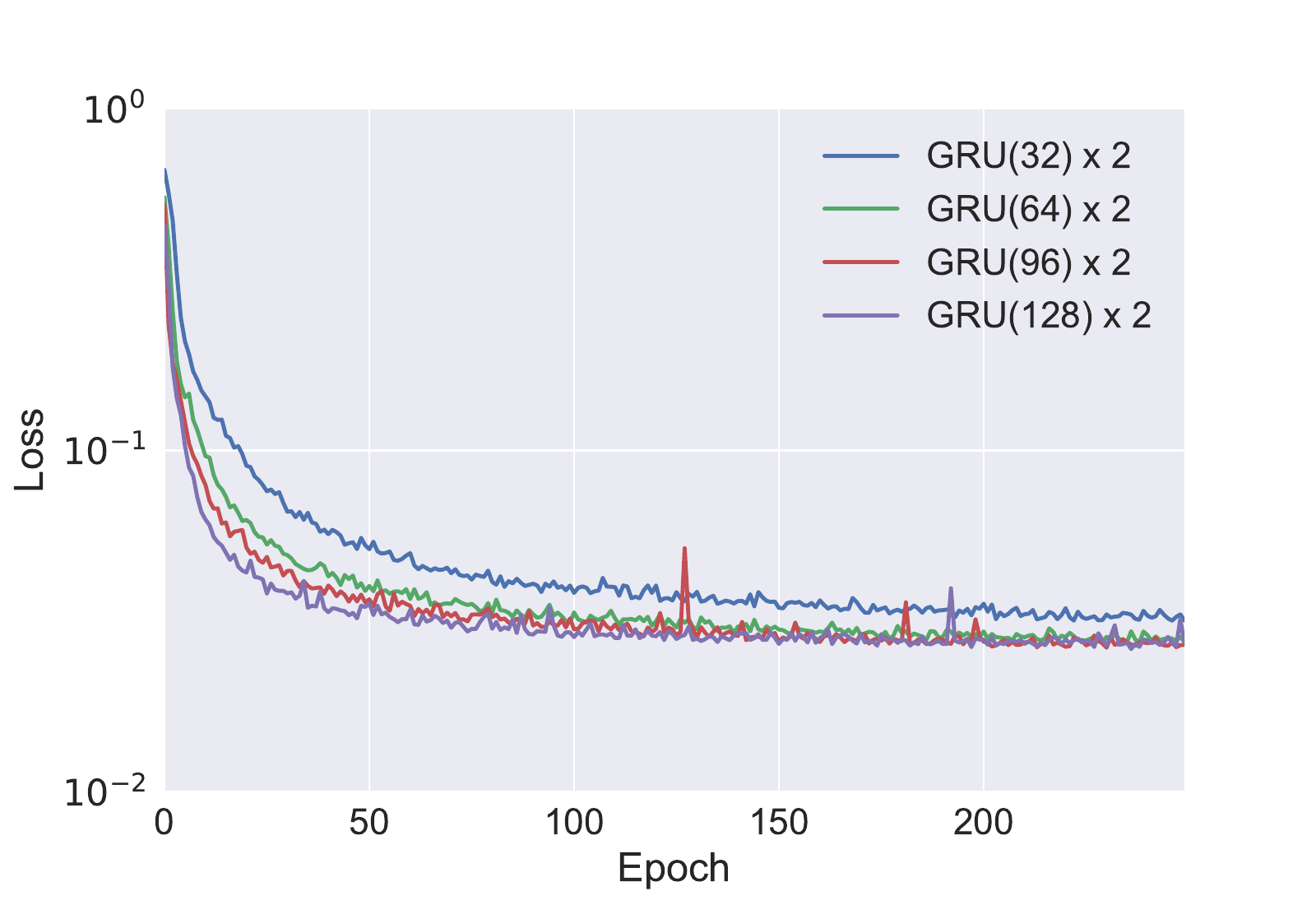}
\caption{By time}
\end{subfigure}
\begin{subfigure}[b]{0.495\textwidth}
\includegraphics[width=\textwidth]{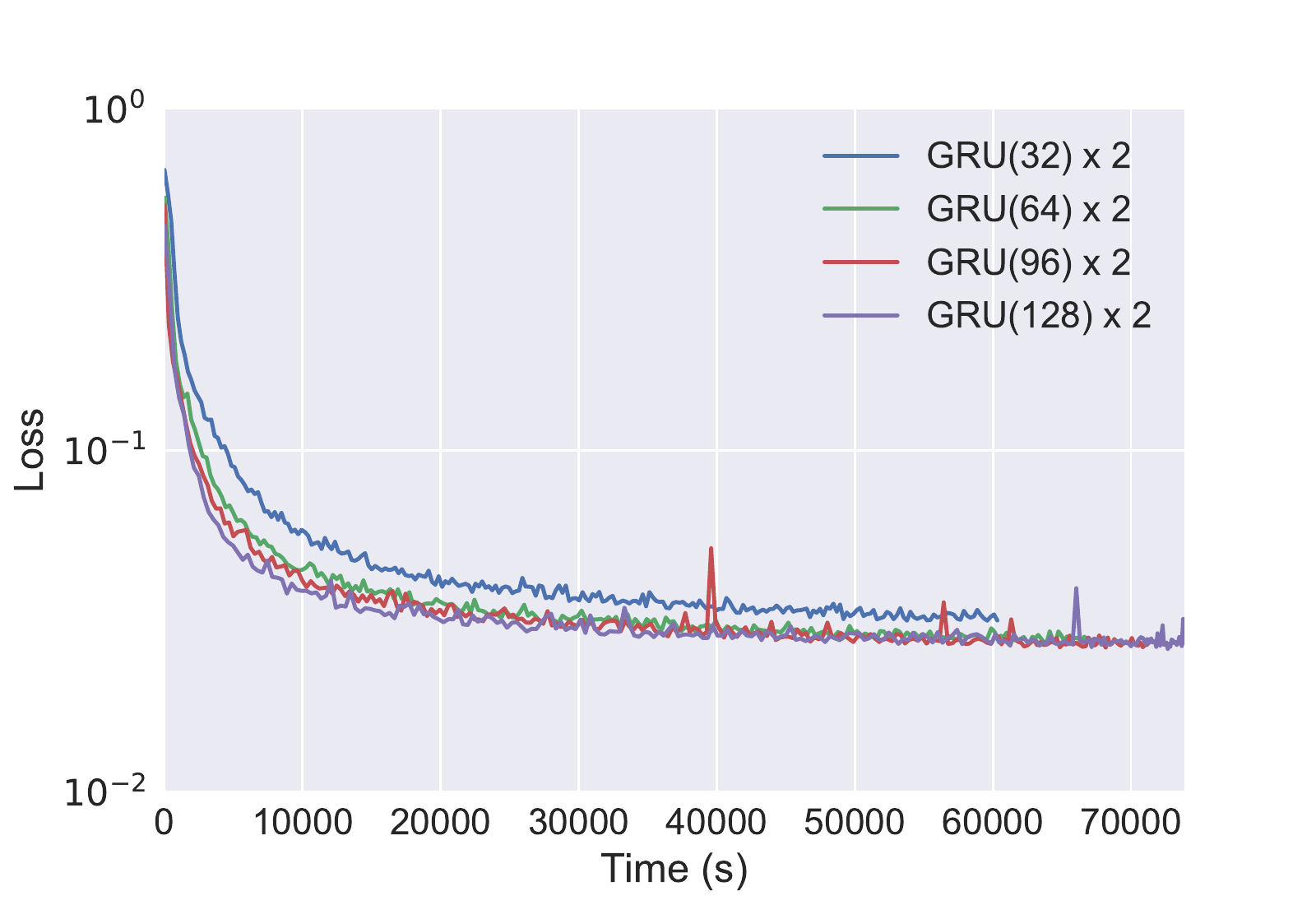}
\caption{By epoch}
\end{subfigure}
\caption{\textbf{Comparison of validation losses during training for various sizes of layers
on the periodic parameters $\bm \theta_i$.} The loss plotted is the mean
squared error over the validation set of the parameters being estimated as a function of
a) training epoch and b) training wall time.  Each colored line corresponds to a specific
network architecture (layer type, layer size, and network depth) as shown in the legend:
for example, ``GRU(64) $\times$ 2'' denotes two decoder GRU layers each composed of 64
hidden units.}
\label{fig:gru_period}
\end{figure}

Supplementary Figure~\ref{fig:linear_noisy} depicts three realizations of the light curve resampling
process described above.
\begin{figure}[htpb!]
    \centering
    \includegraphics[width=0.495\textwidth]{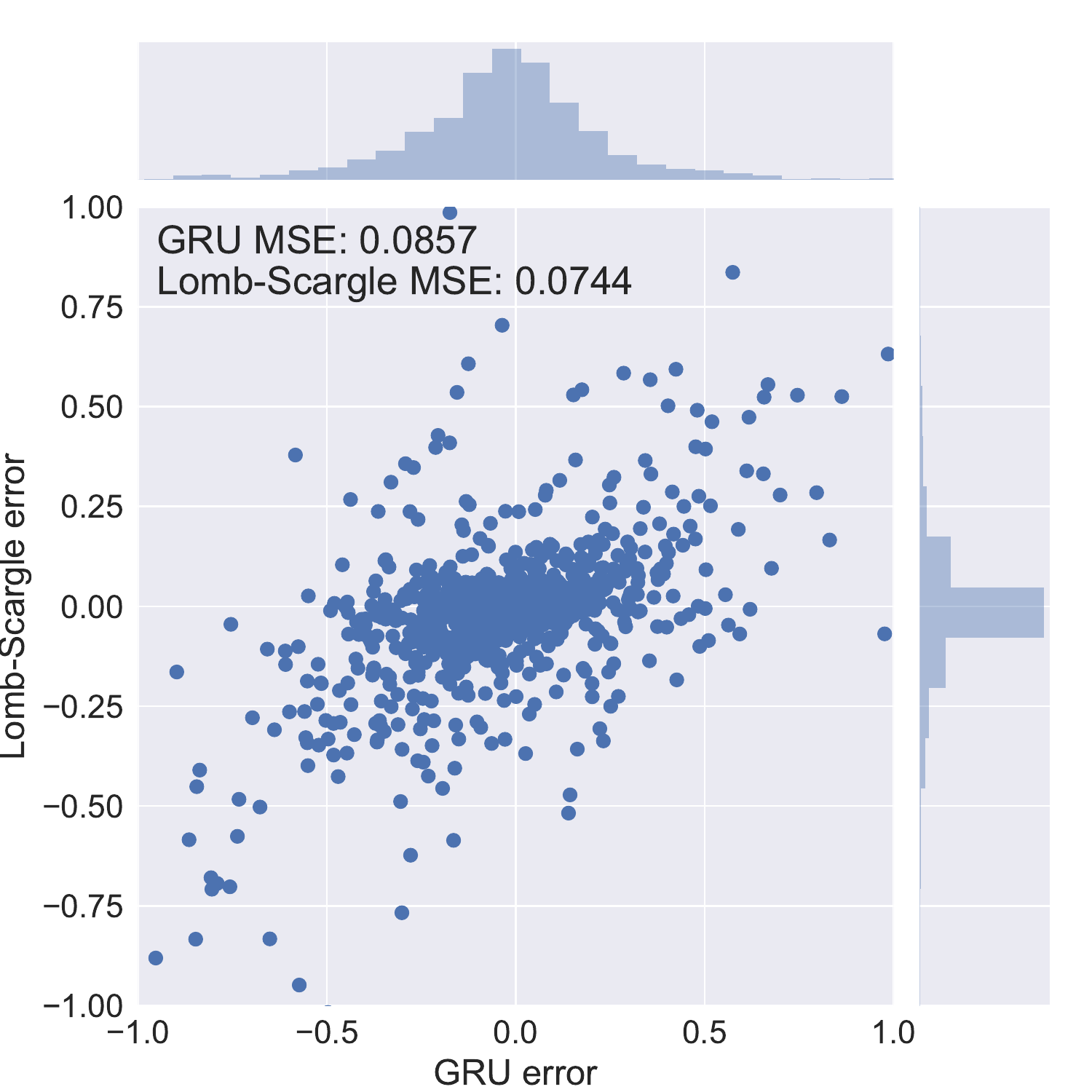}
    \caption{\textbf{Comparison of period estimate residuals $T_i - \hat{T}_i$ for GRU and
Lomb-Scargle models.} Each point on the scatter plot represents the error of the predicted
period from each model; the histograms on the top and right show the overall distributions
of errors across all samples for the GRU and Lomb-Scargle models, respectively.}
    \label{fig:ls_period}
\end{figure}
Supplementary Figure~\ref{fig:ls_period} compares the
accuracy of an RNN estimator with two 96-unit GRU layers (ignoring the other estimated
parameters besides the period) with that of a Lomb-Scargle estimate of the period, which
has well-understood statistical convergence properties~\cite{schwarzenberg1991accuracy}.
The accuracy of the two models is comparable, with the Lomb-Scargle estimate performing
slightly better for this particular case. In the next we estimate
the same periodic parameters using a semi-supervised approach based on autoencoder
features.
\subsection{Inferring periodic parameters from autoencoding features}
For the same simulated $N_{\textrm{train}}$ periodic sequences generated above, we now
train a full encoder-decoder network as depicted in Supplementary Figure~1 of the main text and
attempt to recover the parameters $\omega_i, A_i, \phi_i$, and $b_i$ from the encoding
$\bm v_i$. Instead of minimizing a loss function for the parameters themselves, we now train
the network to minimize the mean squared reconstruction error
\begin{equation}
    \textrm{MSE} = \frac{1}{N_{\textrm{train}} n_T} \sum_{i=1}^{N_{\textrm{train}}}
    \sum_{j=1}^{n_T} \left(f_i(t_i^{(j)}) - \hat{y}_i^{(j)}\right)^2.
\end{equation}
Note that reconstruction target is the original (de-noised) periodic function, not
to the raw measurement values.

\begin{figure}[htpb]
    \centering
    \begin{subfigure}[b]{0.495\textwidth}
        \includegraphics[width=\textwidth]{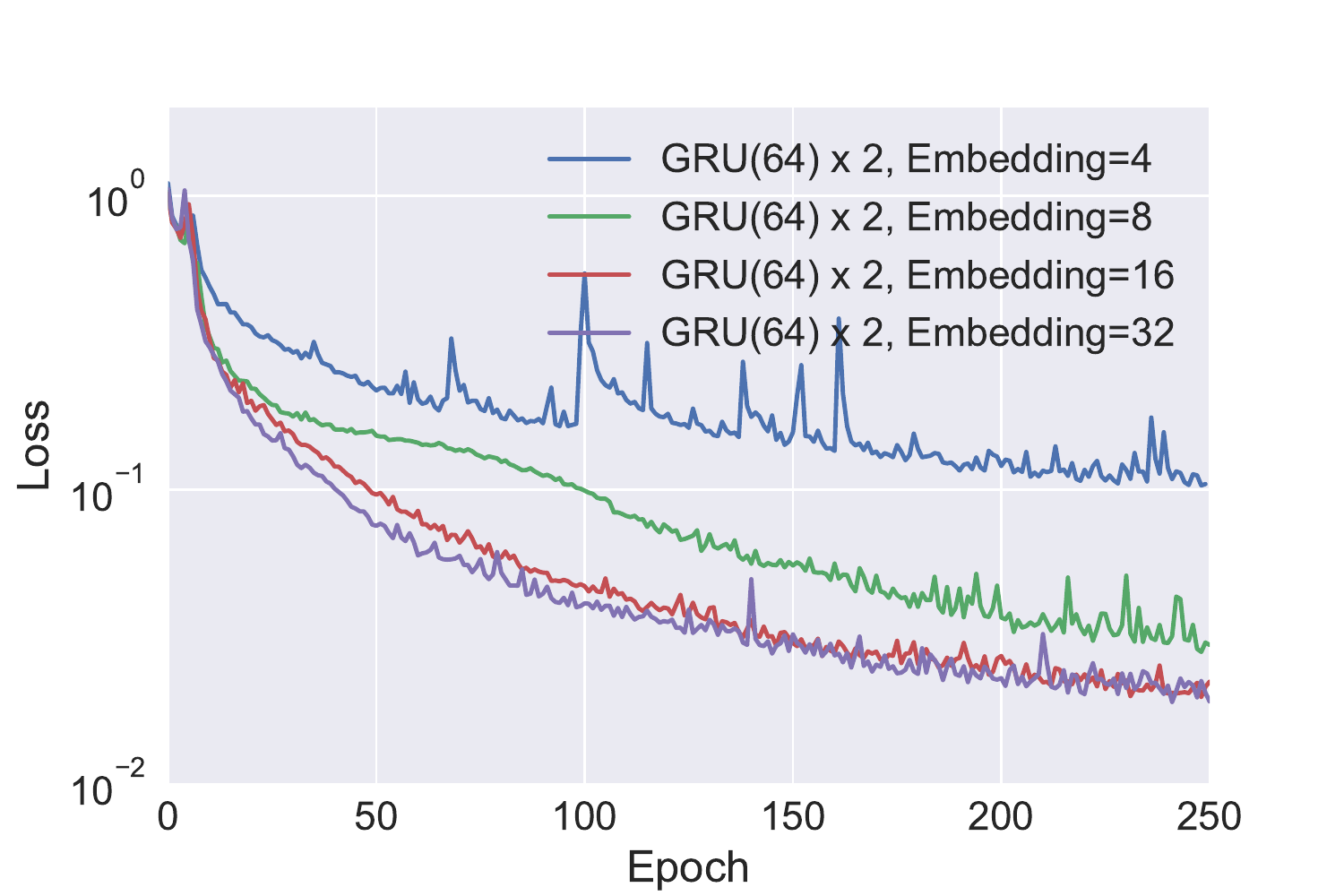}
        \caption{By epoch}
    \end{subfigure}
    \begin{subfigure}[b]{0.495\textwidth}
        \includegraphics[width=\textwidth]{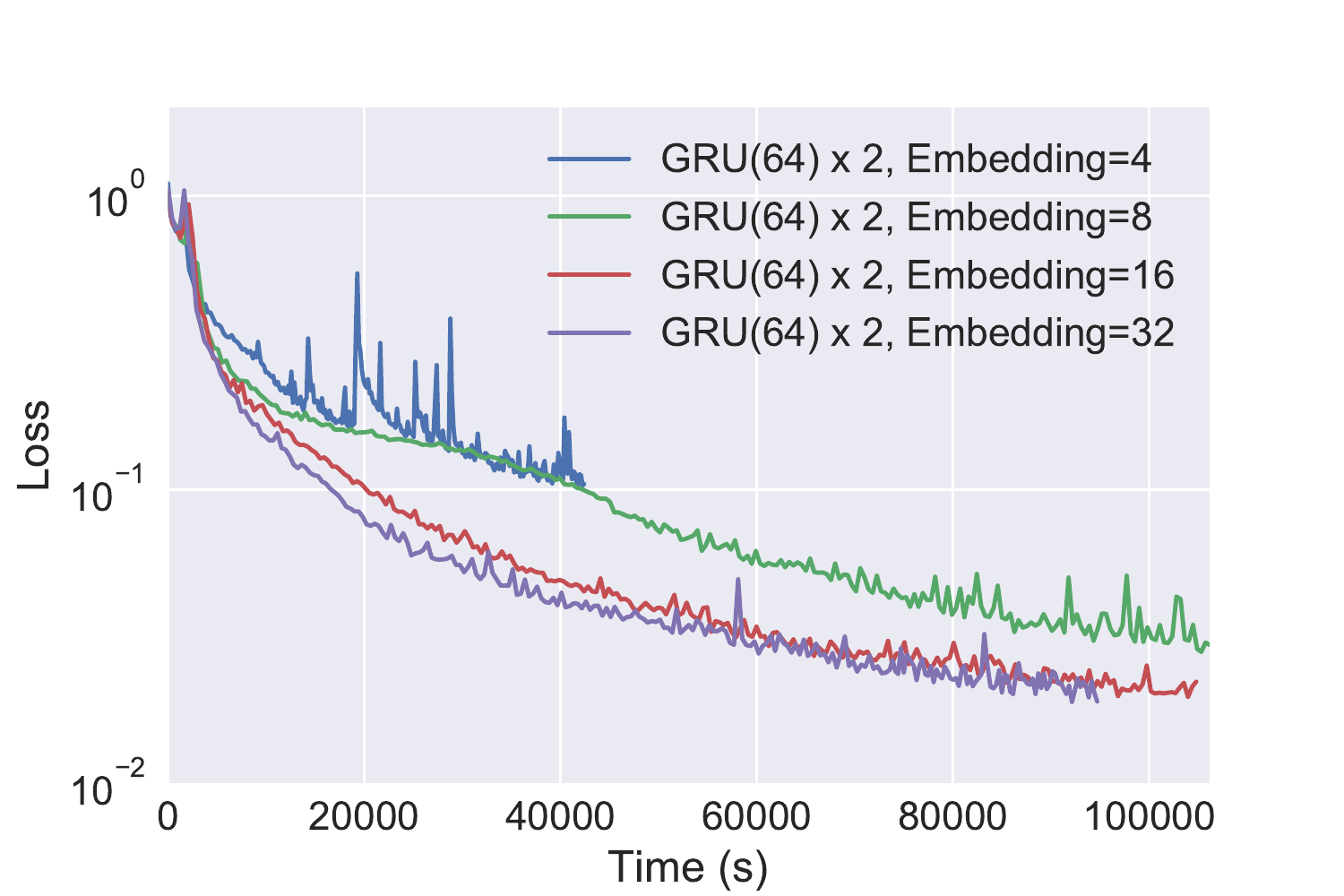}
        \caption{By time}
    \end{subfigure}
    \caption{\textbf{Comparison of GRU autoencoder validation losses during training for various 
    embedding sizes.} The loss
    plotted is the mean squared error over the validation set of the reconstructed sequences 
    as a function of a) training epoch and b) training wall time. Each colored
    line corresponds to a different dimensionality of embedding: for example, ``GRU(64)
    $\times$ 2, Embedding=8'' denotes two encoder and two decoder GRU layers each composed
    of 64 hidden units, with an embedding layer of dimension 8 inbetween.}
    \label{fig:gru_autoencoder}
\end{figure}

The autoencoder model has an additional hyperparameter (compared to the encoder for
directly estimating the periodic parameters) of the size of embedding to use for the
encoding layer. Supplementary Figure~\ref{fig:gru_autoencoder} shows the reconstruction accuracy for a
fixed encoder and decoder architecture, with each consisting of two bidirectional GRU
layers of size 64, for a range of embedding sizes.
Although each target function can be fully characterized in terms of four parameters, it is clear
that increasing the embedding size does improve the accuracy of the reconstruction up to a
point. However, given a longer training time or different choice of learning rate, it is
conceivable that a model with only four embedding values could achieve the same level of
accuracy as those with higher-dimensional representations.

\begin{figure}[htpb!]
    \centering
    \begin{subfigure}[b]{0.495\textwidth}
        \includegraphics[width=\textwidth]{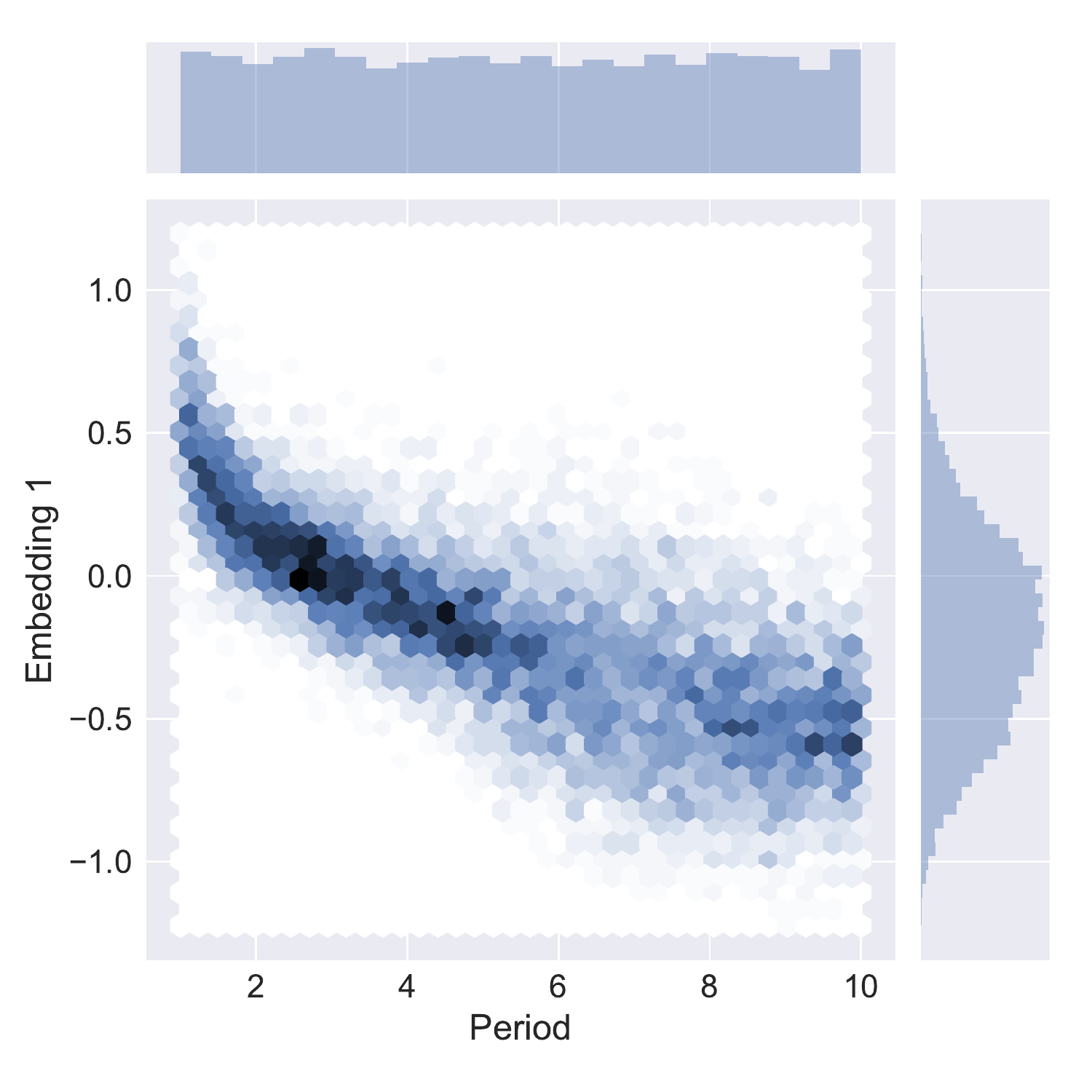}
        \caption{Autoencoder feature vs. period}
    \end{subfigure}
    \begin{subfigure}[b]{0.495\textwidth}
        \includegraphics[width=\textwidth]{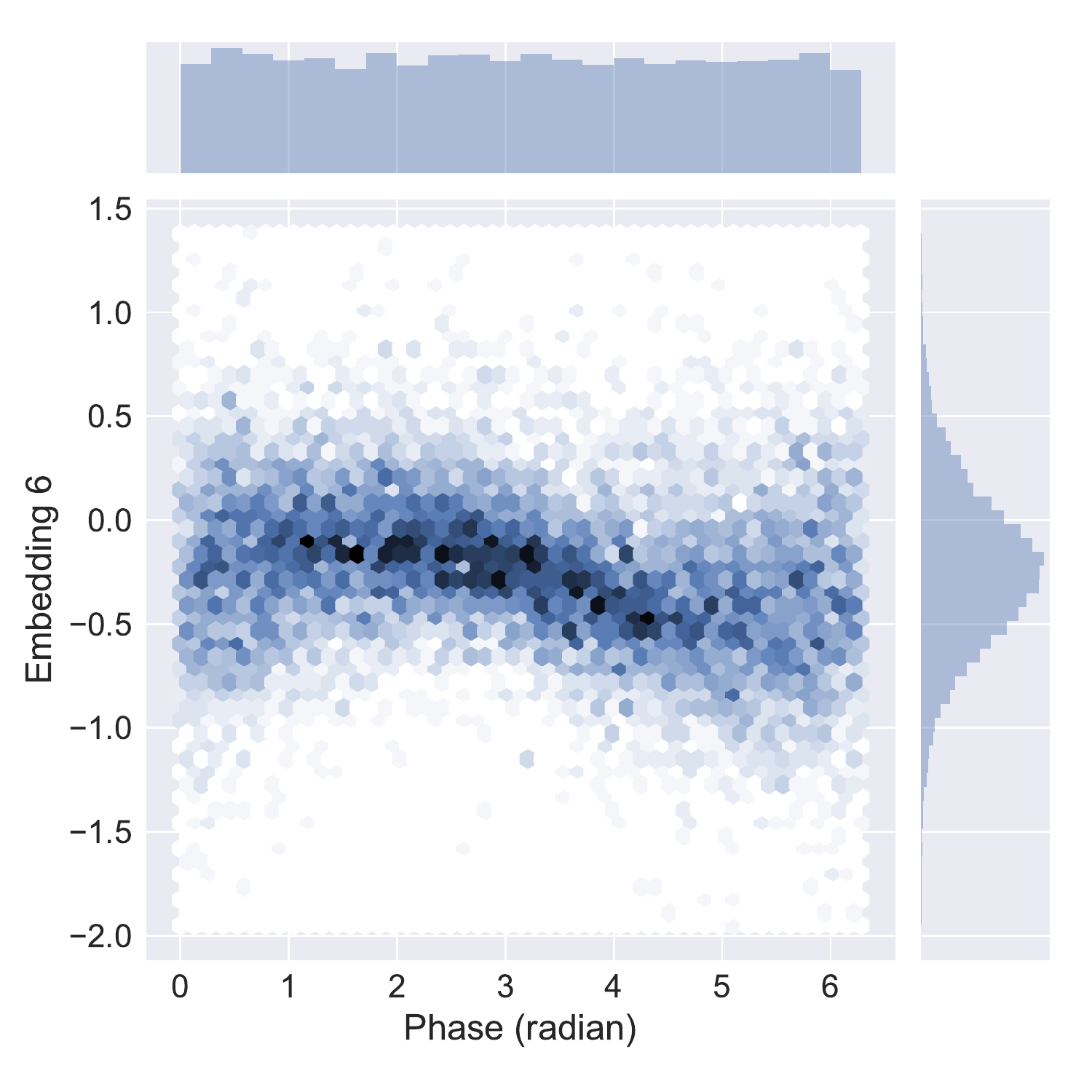}
        \caption{Autoencoder feature vs. phase}
    \end{subfigure}
    \caption{\textbf{Visual depiction of the relationships between 8-dimensional
    autoencoding time series features and period.} We observe clear non-linear
    relationships between some of the learned autoencoder features and the a) period and b)
    phase of the input signals. The darkness of each point represents the relative
    frequency of that region of pairs of values, and the histograms on the top and right
    show the overall distributions of the estimated parameters and relevant autoencoder
    features, respectively.}
    \label{fig:hex}
\end{figure}

\begin{figure}[htpb!]
    \centering
    \includegraphics[width=0.495\textwidth]{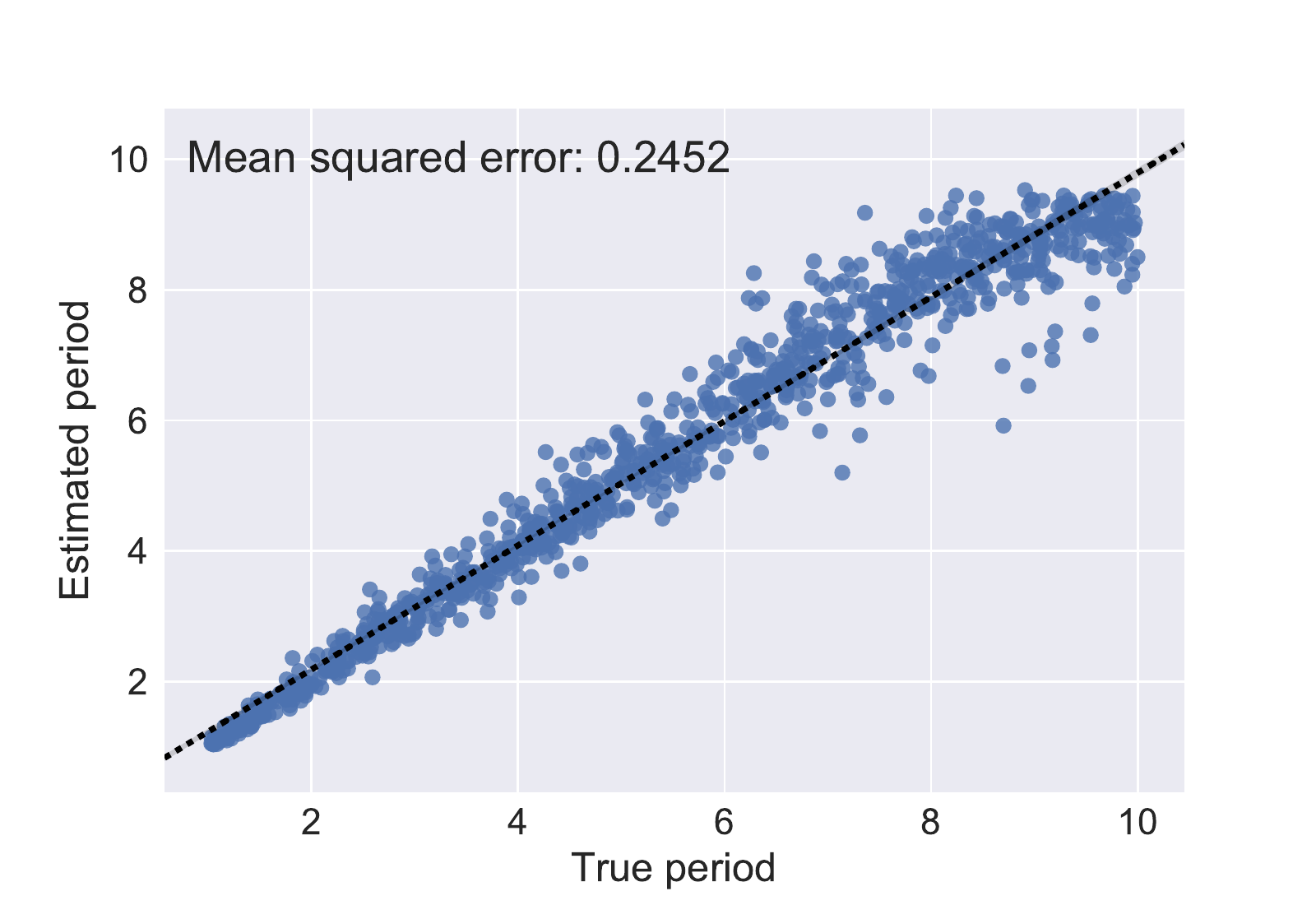}
    \caption{\textbf{True period vs.\ estimated period for random forest regressor model.}
    The dotted black line represents an exact math $T_i = \hat{T}_i$.}
    \label{fig:period_rf}
\end{figure}

Although the inputs to the autoencoder are only the raw time series measurements and not
the periodic parameters, the encoding layer does represents some alternative
fixed-dimensional representation of the entire sequence, so it is reasonable to inquire if
these values can be used to recover the original parameters $\bm \theta_i$.
Supplementary Figure~\ref{fig:hex} demonstrates that strong (non-linear) relationships exist between
some of the encoding values and the period and phase of the input sequences.
While there is not any obvious closed-form relationship that should exist between a single
autoencoding feature and any of the elements of $\bm \theta_i$, we can attempt to train a
separate model mapping the encoding vectors to the parameters of interest.
Since the relationships between the autoencoding features appear to be somewhat
non-linear, we train a random forest regressor~\cite{breiman2001random} with 1000 trees using
\verb scikit-learn ~\cite{pedregosa2011scikit} to estimate the
period, phase, amplitude, and offset of each sequence using the autoencoding features.
Supplementary Figure~\ref{fig:period_rf} shows the accuracy of our model for estimating period from the
encoding values for various representation lengths; the random forest model is trained on
the $N_{\textrm{train}}=50,000$ training sequences and evaluated on the
$N_{\textrm{valid}}=10,000$ validation sequences. The accuracy of the model depends on the
size of the embedding used, as well as indirectly on many other factors:
the size, number, and type of recurrent layers used, the choice of optimization
method and stopping criterion, and on the particular distribution chosen for the simulated
data. Nevertheless, it is clear from this exercise that useful aggregate properties of
time series can be inferred from features automatically generated by an autoencoder. In
the next section we explore how these features can be used within the context of a
real-world supervised classification problem.

\section{Application to astronomical light curves of variable stars}\label{sec:lc}
\subsection{Validation accuracy}
The distinction between training, test, and validation data is a subtle one in our
problem. In order to avoid overly optimistic results, it is crucial to avoid making
use of the test data in any way during the training process. Our experiment, however,
is intended to simulate the case where some labeled and some unlabeled light curves
are available: in this case, the practitioner is free to train the autoencoder using
only the labeled light curves, or using the full dataset. We initially withheld some
validation light curves in order to better understand the difference between
reconstruction accuracy for light curves that are seen during training versus new light
curves; the validation losses reported in Supplementary Figures~\ref{fig:asas_autoencoder}
and~\ref{fig:dropout} are computed using 20\% of the available light curves which were
withheld during training. After confirming that the reconstruction performance for
training and validation light curves was comparable, we re-trained our autoencoder using
all the available data before training our random forest classifiers.

Supplementary Figure~\ref{fig:asas_autoencoder} shows the validation loss over time for various
embedding sizes. As expected, the embedding size required for an accurate reconstruction
is larger than in the case of simulated purely sinusoidal data, and the average reconstruction
error at convergence is somewhat higher. Nevertheless, we find that our autoencoder is able to
accurately model light curves using a relatively parsimonious feature representation.
\begin{figure}[htpb]
    \centering
    \begin{subfigure}[b]{0.495\textwidth}
        \includegraphics[width=\textwidth]{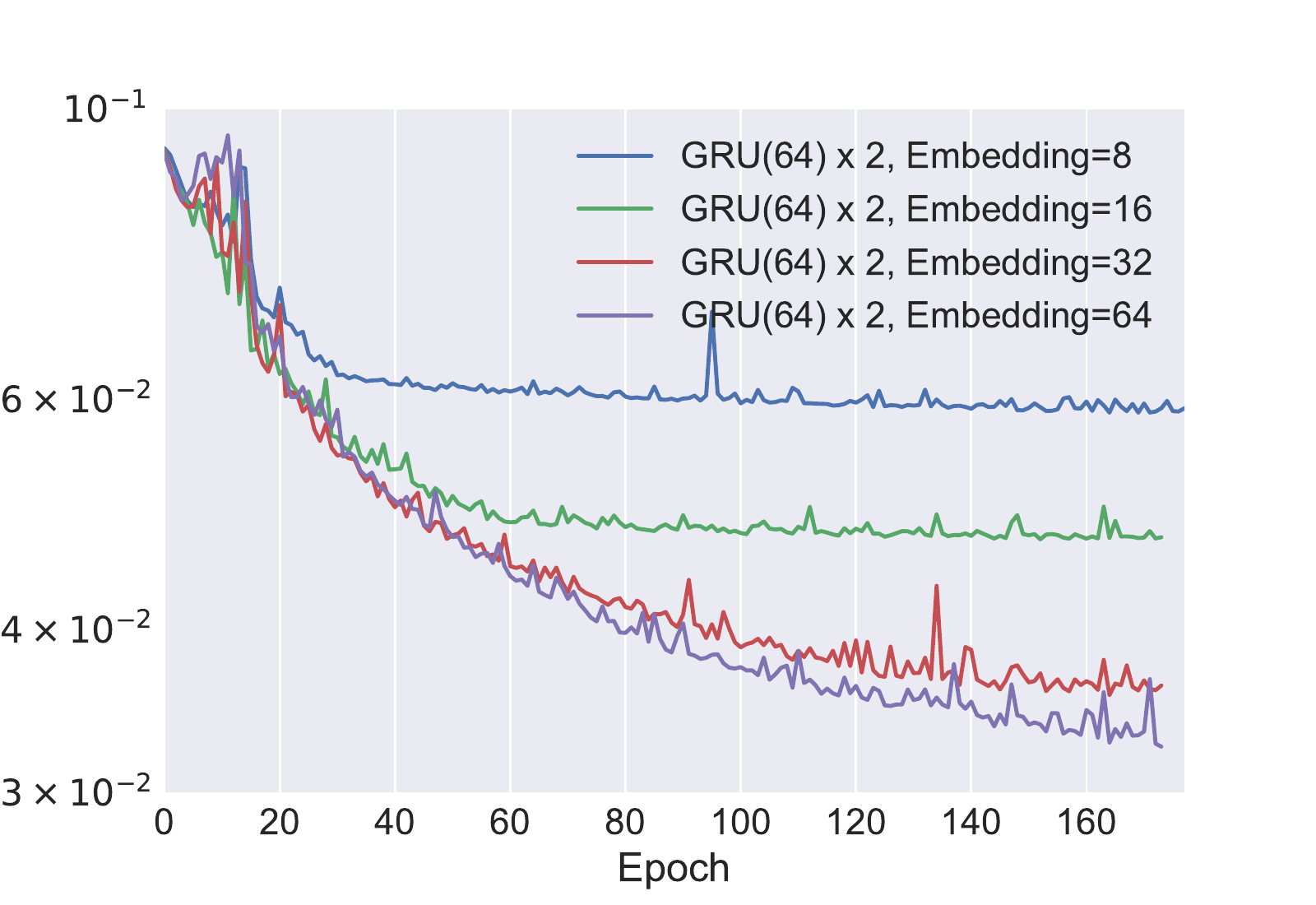}
        \caption{By epoch}
    \end{subfigure}
    \begin{subfigure}[b]{0.495\textwidth}
        \includegraphics[width=\textwidth]{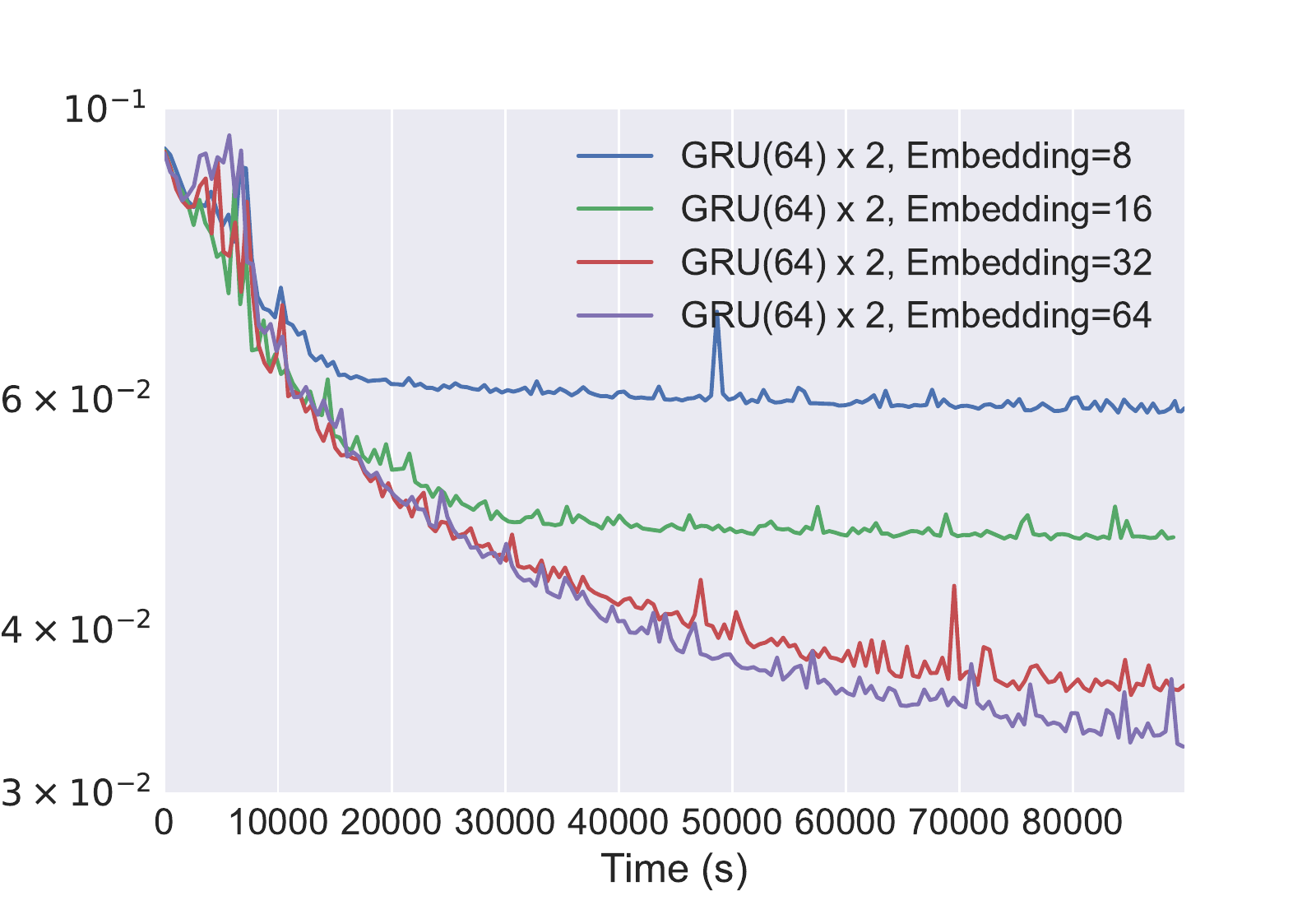}
        \caption{By time}
    \end{subfigure}
\caption{\textbf{Comparison of period-folded light curve autoencoder validation losses during
    training for various embedding sizes.} The loss
    plotted is the mean squared error over the validation set of the reconstructed sequences 
    as a function of a) training epoch and b) training wall time. Each colored
    line corresponds to a different dimensionality of embedding: for example, ``GRU(64)
    $\times$ 2, Embedding=8'' denotes two encoder and two decoder GRU layers each composed
    of 64 hidden units, with an embedding layer of dimension 8 inbetween.}
    \label{fig:asas_autoencoder}
\end{figure}

The above experiments using simulated data show that autoencoders can be used to model
periodic sequence that has been irregularly sampled, and that the resulting encodings
contain information about the frequency domain properties and other global characteristics
of the input time series.

\begin{figure}[htpb!]
    \centering
    \begin{subfigure}[b]{0.495\textwidth}
        \includegraphics[width=\textwidth]{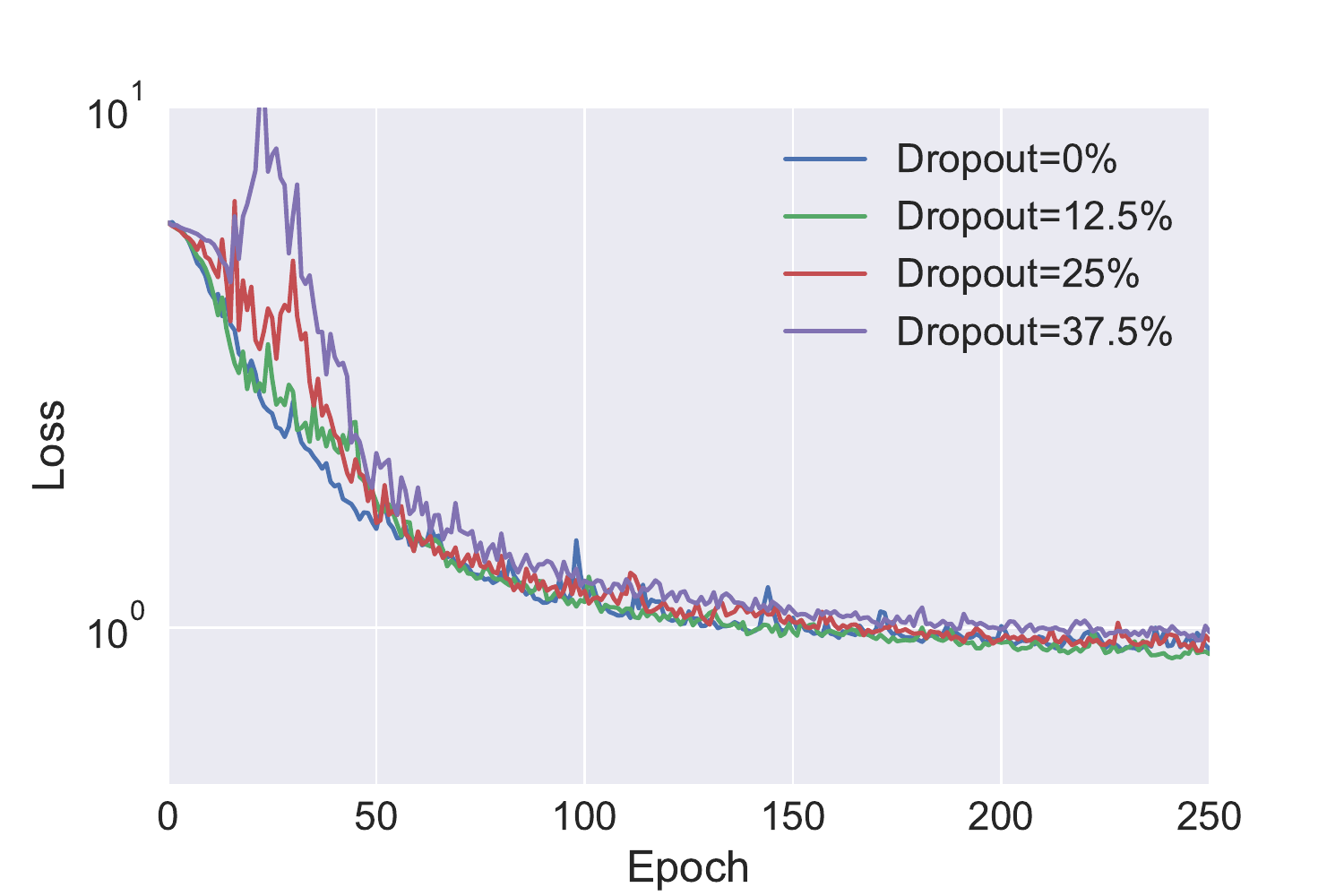}
        \caption{By epoch}
    \end{subfigure}
    \begin{subfigure}[b]{0.495\textwidth}
        \includegraphics[width=\textwidth]{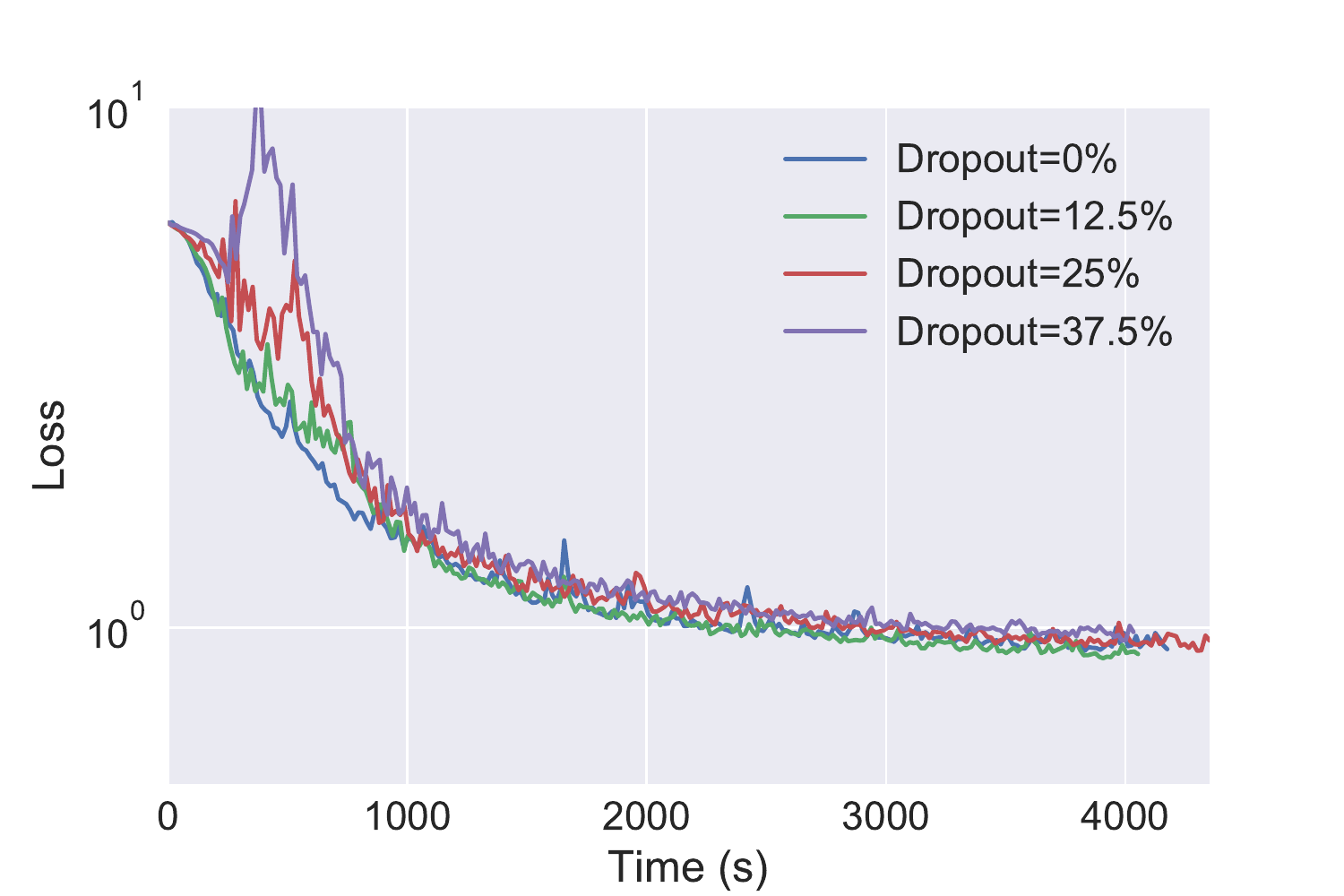}
        \caption{By time}
    \end{subfigure}
\caption{\textbf{Comparison of period-folded light curve autoencoder validation losses during
    training for various levels of dropout.} The loss
    plotted is the mean squared error over the validation set of the reconstructed sequences 
    as a function of a) training epoch and b) training wall time. Each colored
    line corresponds to a different dimensionality of embedding: for example, ``GRU(64)
    $\times$ 2, Embedding=8'' denotes two encoder and two decoder GRU layers each composed
    of 64 hidden units, with an embedding layer of dimension 8 inbetween.}
    \label{fig:dropout}
\end{figure}
As described in Supplementary Figure~1 of the main text, in our experiments we imposed 25\%
dropout~\cite{srivastava2014dropout} between recurrent layers in order to avoid
overfitting. We tested the effect of dropout using the LINEAR dataset since its smaller
size makes overfitting a bigger concern. 
Supplementary Figure~\ref{fig:dropout} shows how varying the dropout parameter affects the resulting
reconstruction validation error; we find that the effect is minimal for this problem.

\subsection{Sequence length}
As described in the main text, before training our autoencoder we split the available
light curves into subsequences of length $n_T=200$. This step is purely optional, since
our RNN architecture can accommodate input and output sequences of arbitrary length.
However, the use of constant-length sequences is advantageous in training: it eliminates
the need to pad input sequences and/or mask output sequences when computing the loss
function, and empirically it seems to improve training speed (as measured in both
iterations and minutes).
\begin{figure}[htpb]
    \centering
    \begin{subfigure}[b]{0.495\textwidth}
        \includegraphics[width=\textwidth]{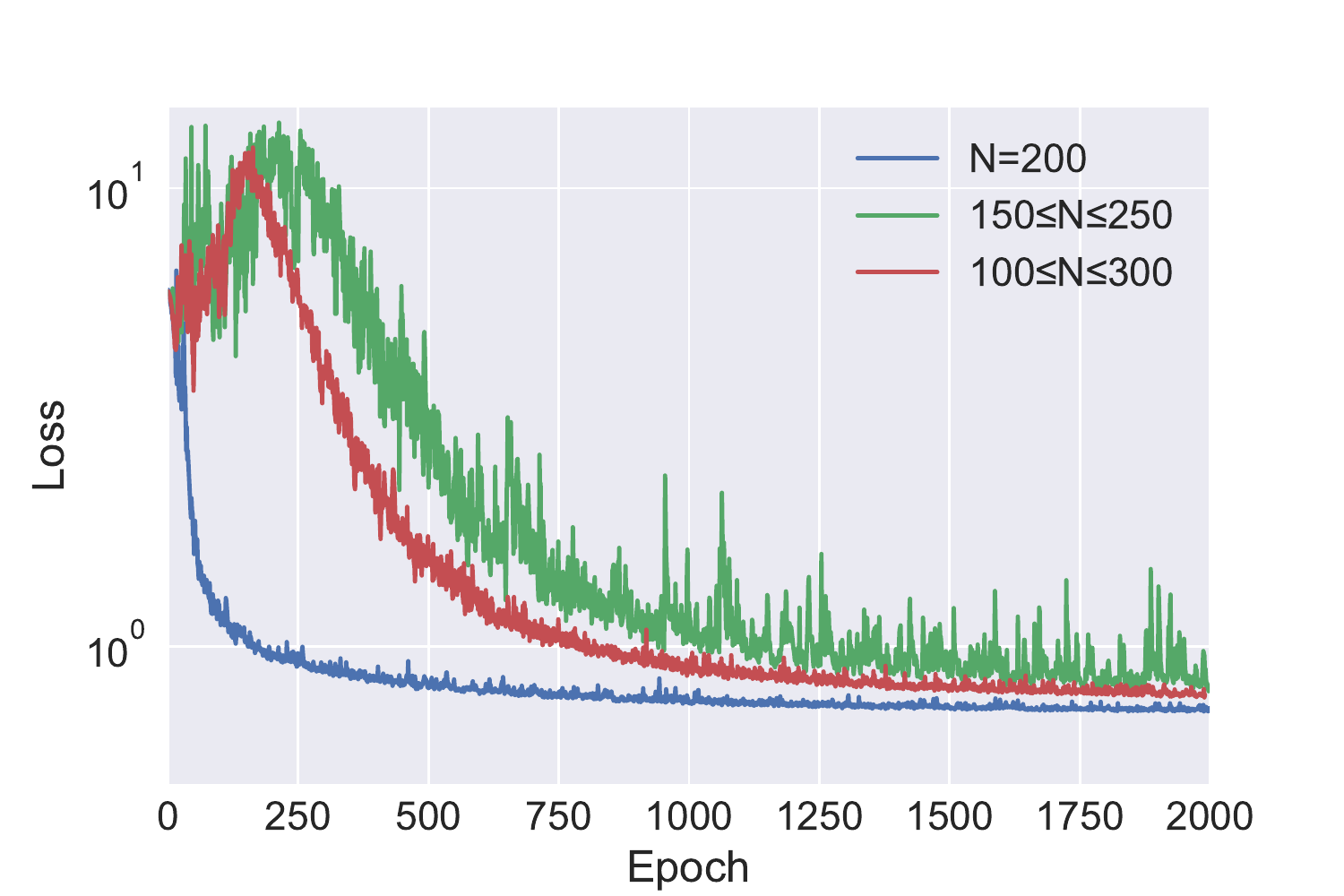}
        \caption{By epoch}
    \end{subfigure}
    \begin{subfigure}[b]{0.495\textwidth}
        \includegraphics[width=\textwidth]{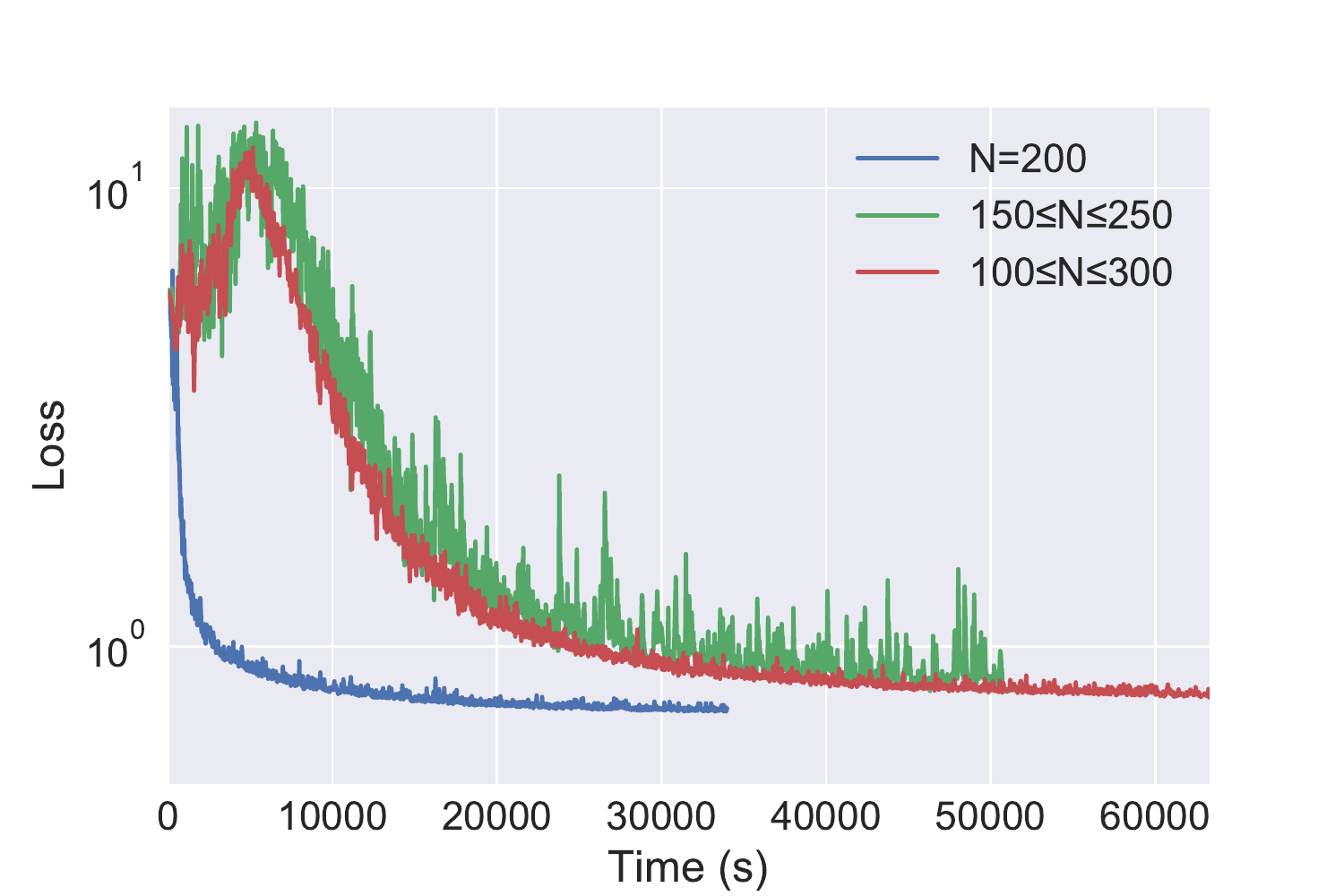}
        \caption{By time}
    \end{subfigure}
\caption{\textbf{Comparison of period-folded light curve autoencoder validation losses during
    training for various sequence lengths.} The loss
    plotted is the mean squared error over the validation set of the reconstructed sequences 
    as a function of a) training epoch and b) training wall time. Each colored
    line corresponds to a different dimensionality of embedding: for example, ``GRU(64)
    $\times$ 2, Embedding=8'' denotes two encoder and two decoder GRU layers each composed
    of 64 hidden units, with an embedding layer of dimension 8 inbetween.}
    \label{fig:var_length}
\end{figure}

Supplementary Figure~\ref{fig:var_length} depicts the autoencoder training process for different lengths
of sequence: first, for constant-size sequences of length $n_T=200$; second, for
variable-length sequences between $150\leq n_T \leq 250$; and finally for $100\leq n_t
\leq 300$. In each case light curves from the LINEAR dataset were used for training, but
for the variable-length sequences the length $n_T$ was chosen at random for each light
curve subsequence. Although the required training time is significantly longer for the
variable-length problem, roughly the same quality of reconstruction is ultimately
achieved. However, it is clear that at some point the difference in sequence length would
become problematic, and another training approach might need to be explored (for example,
grouping different lengths of light curves together and training in batches, or training
different models for different sizes of data).

\subsection{Data augmentation using noise properties}
Data augmentation is commonly applied in the context of image processing tasks in order to
increase the effective volumes of available training data (see, e.g.,~\cite{simard2003best}).
Unlike in the case of image data, for astronomical surveys the noise properties of the
measurements are often known, which provides a natural way to generate synthetic training
examples that mimic additional samples from the same survey for a given source.
In particular, we attempted to augment our light curve datasets by generating new training
samples from existing light curves by adding Gaussian noise corresponding to the estimated
measurement error at each time step:
\begin{equation}
    \bm \tilde{y}_i = \left(y_i^{(1)} + \epsilon_i^{(1)}, \dots, y_i^{(n_T)} +
    \epsilon_i^{(n_T)}\right),\ \epsilon_i^{(j)} \sim \mathcal{N}(0, \sigma_i^{(j)}).
\end{equation}
\begin{figure}[htpb!]
    \centering
    \includegraphics[width=0.495\textwidth]{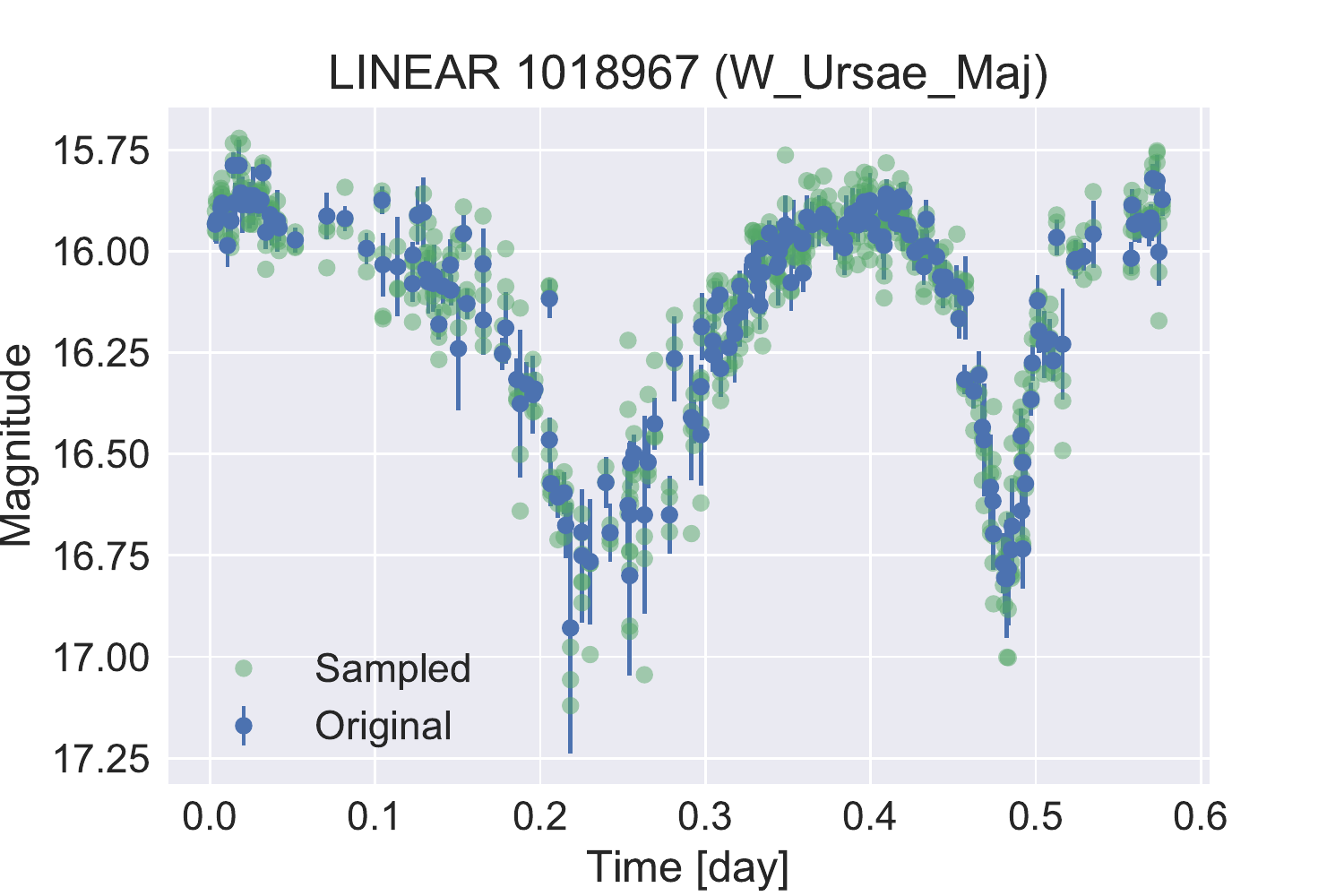}
    \caption{Examples of generated LINEAR light curves using noise properties}
    \label{fig:linear_noisy}
\end{figure}
Supplementary Figure~\ref{fig:linear_noisy} depicts three realizations of the light curve resampling
process described above.

Training our autoencoder on the LINEAR dataset for the same number of epochs but using the
additional training samples leads to an increase in validation accuracy from 72.6\% to
78.1\% for the non-period folded model. However, for the period-folded model, the use of
additional training examples did not improve performance, and in fact slightly diminished
the validation accuracy (from 97.1\% to 96.7\%); this is likely because the autoencoder
model is already able to accurately reconstruct the period-folded light curves, whereas
the raw light curves are not well-modeled and therefore the additional training data is
useful. We also attempted applying the same type of noise resampling for the previous
example using the ASAS dataset, but the change in accuracy for the resulting classifier
was negligible.

\subsection{Direct supervised classification}
The architecture we have described can easily be modified for
other tasks, including direct supervised classification of unevenly sampled time series:
in this case, the decoder module in Figure~1 would be
replaced by one or more fully-connected layers, and the network would be trained to
minimize the multi-class cross entropy classification loss. Such an approach trained on
the period-folded data and sampling times also leads to high classification accuracy,
achieving 98.6\% average validation accuracy for the ASAS dataset and 96.7\% for the
LINEAR dataset. However, the fully supervised approach requires a large amount of labeled
training data, whereas the autoencoder method can make use of either labeled or unlabeled
data; the use of unsupervised methods is common in such cases where limited labeled data
is available~\cite{glorot2011domain}.

\subsection{Transfer learning}
In the experiments in the main text, data from a single survey is used to train an autoencoder,
which is then used to produce features used for classification of labeled training
examples from the same survey. Because the autoencoder and random forest models are
decoupled, it is straightforward to incorporate additional or completely different
training data for the unsupervised portion of the model.  In the ASAS example, we make use
of both labeled and unlabeled examples when training the autoencoder; however, we could
just as easily train the autoencoder using only unlabeled data, or data from an entirely
different source. As a simple example, we found that features from the autoencoder trained
on ASAS or LINEAR data yielded nearly the same level of classification accuracy for light
curves from the other survey (less than a 1\% reduction in each case).

\end{document}